\documentclass[preprint,preprintnumbers,nofootinbib,aps,twocolumn,10pt]{revtex4-1}
\usepackage{amsmath,amssymb,bm,epsfig}
\usepackage{color}
\usepackage{natbib}
\usepackage{hyperref} 
\usepackage{ulem}
\usepackage{graphicx}
\RequirePackage{lineno}

\usepackage{xcolor,colortbl}
\usepackage{pifont}

\def \beq{\begin{equation}}
\def \eeq{\end{equation}}
\def \beqa{\begin{eqnarray}}
\def \eeqa{\end{eqnarray}}
\def \la{\langle}
\def \ra{\rangle}
\def \l{\left(}
\def \r{\right)}

\usepackage{xspace}

\newcommand{\sNN}{\sqrt{s_{\rm NN}}}

\begin{document}
\title{Initial conditions from the shadowed Glauber model for Pb+Pb at 
$\sqrt{s_{\rm NN}}=2.76$ TeV}

\author{Snigdha Ghosh}
\email{snigdha.physics@gmail.com}
\affiliation{Theoretical Physics Division, 
Variable Energy Cyclotron Centre, 1/AF Bidhannagar, 
Kolkata, 700064, India}

\author{Sushant K. Singh}
\email{sushantsk@vecc.gov.in}
\affiliation{Theoretical Physics Division, 
Variable Energy Cyclotron Centre, 1/AF Bidhannagar, 
Kolkata, 700064, India}

\author{Sandeep Chatterjee}
\email{sandeepc@vecc.gov.in}
\affiliation{Theoretical Physics Division, 
Variable Energy Cyclotron Centre, 1/AF Bidhannagar, 
Kolkata, 700064, India}

\author{Jane Alam}
\affiliation{Theoretical Physics Division, 
Variable Energy Cyclotron Centre, 1/AF Bidhannagar, 
Kolkata, 700064, India}

\author{Sourav Sarkar}
\affiliation{Theoretical Physics Division, 
Variable Energy Cyclotron Centre, 1/AF Bidhannagar, 
Kolkata, 700064, India}

\begin{abstract}
We study the initial conditions for Pb+Pb collisions at $\sqrt{s_{\rm NN}}=2.76$ TeV 
using the two component Monte-Carlo Glauber model with shadowing of the  nucleons 
in the interior by the  leading  ones. The model parameters are fixed by comparing to 
the multiplicity data of p+Pb and Pb+Pb at $\sqrt{s_{\rm NN}}=5.02$ and $2.76$ TeV 
respectively. We then compute the centrality dependence of the eccentricities upto the 
fourth order as well as their event by event distributions. 
The inclusion of shadowing brings  the Monte-Carlo Glauber model predictions in 
agreement with data as well as with results from other dynamical models of initial 
conditions based on gluon saturation at high energy nuclear collisions. Further, 
we find that the shadowed Glauber model provides the desired relative magnitude between 
the ellipticity and triangularity of the initial energy distribution required to explain 
the data on the even and odd flow harmonics $v_2$ and $v_3$ respectively at the LHC.
\end{abstract}
\maketitle

\section{Introduction}\label{sec.intro}
One of the most important  ingredients  in understanding the evolution of matter 
formed in heavy ion collisions (HIC) is its initial condition (IC). Currently there are 
several models of IC available with varying degrees of success in explaining the 
data~\cite{Bialas:1976ed,Eskola:1999fc,Kharzeev:2000ph,Kharzeev:2001yq,Hirano:2005xf,
Miller:2007ri,Broniowski:2007nz,Schenke:2012wb,Gale:2012rq,Rybczynski:2012av,
Goldschmidt:2015kpa,Niemi:2015qia}. The Monte-Carlo based IC models generate the event 
by event (E/E) fluctuations in observables which can be compared to those measured in 
experiments. Most of these models share the first step- sampling the positions of the 
constituent nucleons of the two colliding nuclei from their nuclear density distribution 
which is usually taken to be a Woods-Saxon profile~\cite{Woods:1954zz}. In the second step, 
they all differ in the energy deposition scheme corresponding to a specific configuration 
of the nucleon positions. This finally results in different predictions of centrality 
dependence of multiplicity, eccentricities and their event by event distributions. The 
largest source of uncertainties on the extracted values of the medium properties obtained 
by comparing the predictions of the theoretical models to data is  known to stem from the 
choice of the IC~\cite{Song:2010mg}.
 
Monte-Carlo Glauber models (MCGMs) have been reasonably successful in describing the qualitative 
features of various observables~\cite{Miller:2007ri,Broniowski:2007nz}. The energy deposition 
scheme is largely geometrical with the only dynamical input being a constant nucleon-nucleon 
cross-section $\sigma_{NN}$. The recent data on $v_2-\frac{dN_{ch}}{d\eta}$ correlation for 
the top ZDC events in U+U collisions at $\sNN=193$ GeV and Au+Au interactions at $\sNN=200$ GeV 
could  not be reproduced within the ambit of the standard  MCGM~\cite{Adamczyk:2015obl,Goldschmidt:2015kpa}. 
The MCGM predictions are also in disagreement with the E/E distribution of the second flow 
harmonic for Pb+Pb collisions at $\sNN=2.76$ TeV. However, dynamical models based on gluon 
saturation physics like IP-Glasma~\cite{Gale:2012rq} and EKRT~\cite{Niemi:2015qia} are in 
agreement with data. 

MCGMs provide a simple and intuitive description of the IC and hence there have been considerable 
efforts to address the above issues within the geometric approach of the 
MCGM~\cite{Rybczynski:2012av,Moreland:2014oya}. Recently, 
we have shown that the inclusion of shadowing effect due to the leading nucleons on those located 
deep inside provides a simple and physical picture that brings the predictions of the shadowed 
MCGM (shMCGM) in agreement with that of data as well as dynamical models like IP-Glasma at 
the top RHIC energy for Au+Au as well as U+U collisions~\cite{Chatterjee:2015aja}. In this paper 
we present the results of the shMCGM for Pb+Pb collisions at $\sNN=2.76$ TeV and compare with 
IP-Glasma model predictions~\cite{Gale:2012rq} as well as the LHC data~\cite{Aamodt:2010cz,
Timmins:2013hq,Aad:2013xma,Aad:2014vba}.

In the next section~\ref{sec.model}, we provide the details of our shMCGM and the values of the 
parameters of the model estimated by comparing with data. In section~\ref{sec.results} we present 
the results obtained in the shMCGM. We provide estimates for the centrality dependence of various 
eccentricities and their E/E distributions. We find good quantitative agreement with data as well 
as IP-Glasma results. Finally, in section~\ref{sec.summary} we summarise.

\section{The Model}\label{sec.model}

The details of the shMCGM are given in Ref.~\cite{Chatterjee:2015aja}. Here we summarise its main 
features. The shMCGM is an extension of the two component MCGM. In the latter, the energy deposited 
at $\l x,y\r$ on the plane transverse to the beam axis (which is along the z axis) is assumed to be 
a linear superposition of two terms- $N_{part}\l x,y\r$ and $N_{coll}\l x,y\r$ where $N_{part}$ 
counts the number of participant nucleons and $N_{coll}$ is the number of possible binary collisions 
between them. The total charged multiplicity $dN_{ch}/d\eta$ is also assumed to have a similar linear 
relation with the total $N_{part}$ and $N_{coll}$
\beqa
\epsilon\l x,y\r &=& \epsilon_0\left[ \l1-f\r N_{part}\l x,y\r + fN_{coll}\l x,y\r\right]\\
\frac{dN_{ch}}{d\eta} &=& \left[ \l\frac{1-f}{2}\r\sum_i^{N_{part}}w_i + f\sum_i^{N_{coll}}w_i\right]
\eeqa
where the $w_i$s are sampled from a negative binomial distribution ($P_{NBD}\l w,n_0,k\r$) with 
variance$\sim\frac{1}{k}$~\cite{Bozek:2013uha}
\beq
P_{NBD}\l w,n_0,k\r = \frac{\Gamma\l k+w\r}{\Gamma\l k\r\Gamma\l w+1\r}\frac{n_0^wk^k}{\l n_0+k\r^{w+k}}
\eeq
$\epsilon_0$ and $n_0$ are the overall normalization parameters for the energy deposited and 
multiplicity produced respectively. $f$ is usually called the hardness factor 
which is fixed by comparing with data. The criterion for a binary collision between nucleon $i$ 
from nucleus $A$ at $\l x_A^i,y_A^i\r$ and nucleon $j$ from nucleus $B$ at $\l x_B^j,y_B^j\r$ is given 
by $r_{AB}^{ij}\leq \frac{\sigma_{NN}}{\pi}$ where $r_{AB}^{ij}$ is the squared distance in the 
transverse plane between the two nucleons
\beq
r_{AB}^{ij} = \l x^i_A-x^j_B\r^2 + \l y^i_A-y^j_B\r^2\label{eq.collcriteria}
\eeq
All nucleons that suffer at least a single binary collision is treated as a participant. Thus in the 
standard two component MCGM approach all the participants are treated democratically irrespective of 
their positions along $z$-axis. In the shMCGM we introduce the effect of shadowing due to leading participants 
on the others through the following ansatz
\beq
S\l n,\lambda\r = e^{- n\lambda}\label{eq.shadow}
\eeq
where $S\l n,\lambda\r$ is the shadowing effect on a participant due to $n$ other nucleons from the 
same nucleus which are in front and shadow it. Thus all the participants are no more treated on equal 
footing- the leading nucleons contribute to energy deposition more than those located deep inside. Thus 
overall we have the following four parameters in the shMCGM- $n_0$, $k$, $f$ and $\lambda$ which are 
constrained by data.

\section{Results}\label{sec.results}

\begin{figure}[]
\begin{center}
\includegraphics[angle=-90, scale = 0.32]{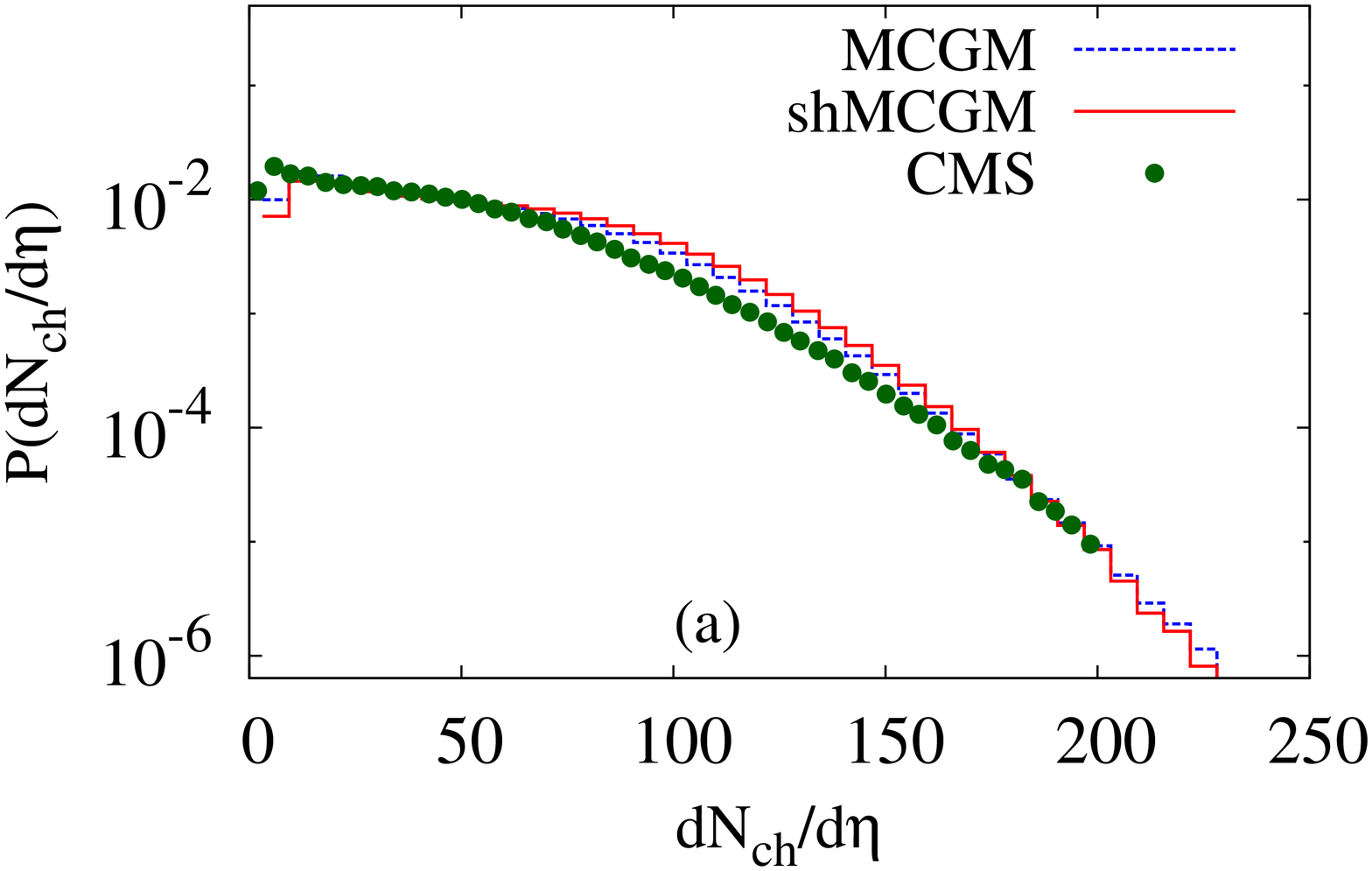}
\includegraphics[angle=-90, scale = 0.32]{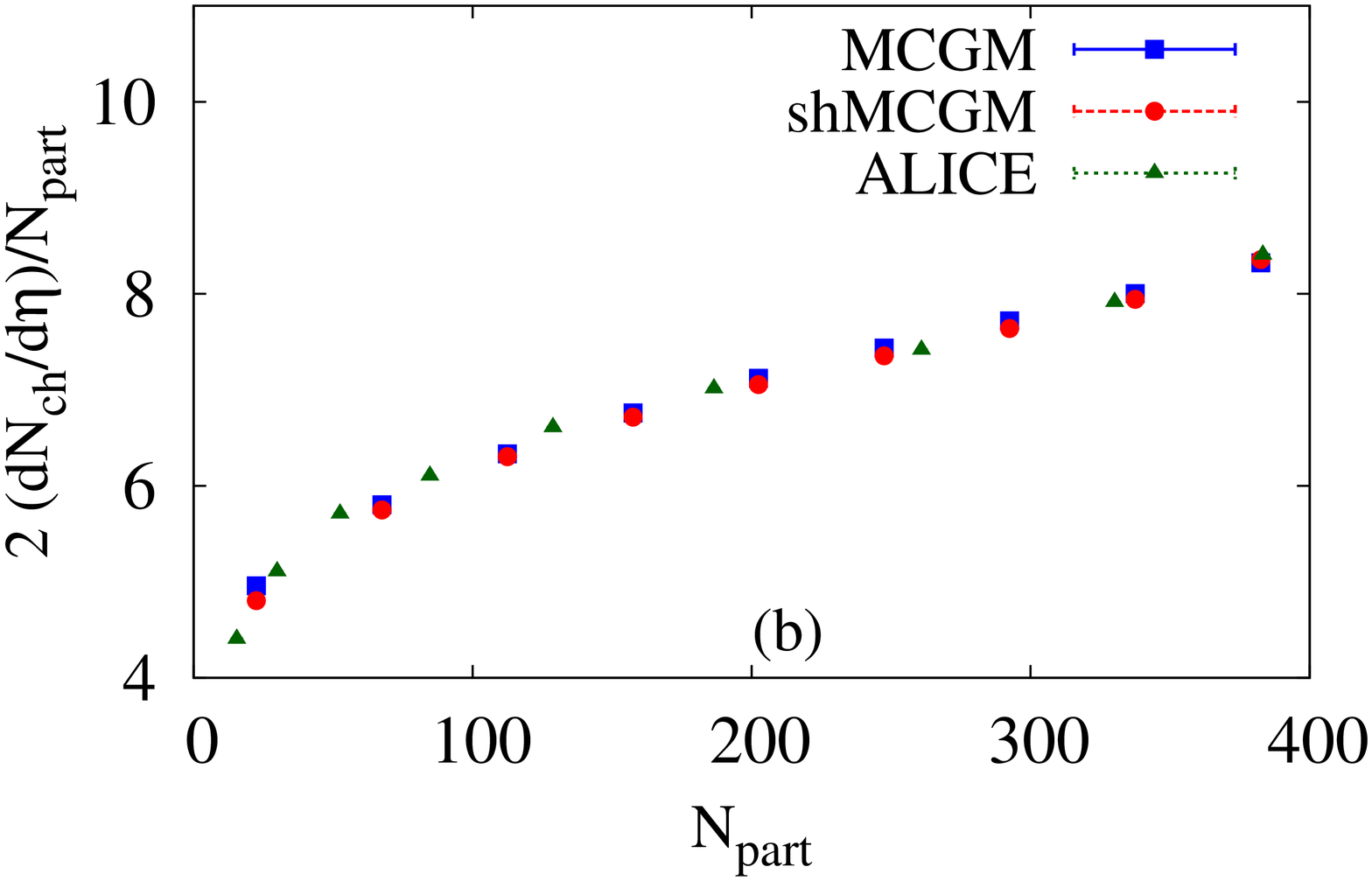}
\end{center}
\caption{(Color online) 
{\it Top}: The probability distribution of $dN_{ch}/d\eta$ for p+Pb at 
$\sNN=5.02$ TeV compared between data~\cite{CMS:2012}, MCGM and shMCGM. {\it Bottom}: The centrality 
dependence of $dN_{ch}/d\eta$ for Pb+Pb at $\sNN=2.76$ TeV compared between data~\cite{Aamodt:2010cz}, 
MCGM and shMCGM.
}
\label{fig.dnchdy}\end{figure}
We will now present the results of the shMCGM. In Fig.~\ref{fig.dnchdy} (a) we have plotted the 
probability distribution of charged particle multiplicity observed in p+Pb collision at 5.02 TeV for 
$|\eta|<2.5$~\cite{CMS:2012}. We have compared the data with results from both MCGM as well as shMCGM. 
The nucleons are sampled from a Woods-Saxon distribution of the Pb nuclear density. The Woods-Saxon 
parameters of the Pb nucleus are: the radius $R=6.7$ fm obtained from the parametrisation 
$R=1.12A^{0.33}-0.86A^{-0.33}$ and the surface diffusion $\delta=0.54$ fm~\cite{Kolb:2003dz}. We find 
good description of the long tail in the distribution only after including NBD fluctuation~\cite{Bozek:2013uha} 
with $k\sim1$. In Fig.~\ref{fig.dnchdy} (b) we display the centrality dependence of the charged particle 
multiplicity in Pb+Pb collisions at $\sNN=2.76$ TeV and contrast the results with the available data~\cite{Aamodt:2010cz}.
It is clear that this plot does not discriminate between MCGM and shMCGM as both the models describe 
the data well for suitably adjusted values of the model parameters. Thus the shadow parameter $\lambda$ 
can not be fixed from this plot. We have fixed $\lambda$ from the E/E distribution plot of scaled $v_2$ 
assuming linear hydrodynamic response, i.e. $v_2\propto\varepsilon_2$. The values of the parameters so obtained 
is tabulated in  Table~\ref{tab.params}. Apart from the parameters listed in Table~\ref{tab.params}, 
we need  the nucleon-nucleon cross-section $\sigma_{NN}$ which is taken here as $64$ mb (the 
corresponding value at RHIC is $42$ mb). We take $\sigma=0.5$ fm, which is used in the Gaussian ansatz 
below to smear the energy $\epsilon_i$ deposited by a participant located at $\l x_i, y_i\r$,
\beqa
\epsilon_i\l x,y\r &=& \frac{\epsilon_0}{2\pi\sigma^2}e^{-\frac{\l x-x_i\r^2+\l y-y_i\r^2}
{2\sigma^2}}
\label{eq.gaussiansource}
\eeqa

\begin{table}[b]
\begin{center}
\begin{tabular}{|r|c|c|c|c|}
\hline
Model & $n_0$ & $f$ & $k$ & $\lambda$ \\
\hline
MCGM & 4.05  & 0.11 & 1 & -\\
\hline
shMCGM & 4.05 & 0.32 & 1 & 0.08 \\
\hline
\end{tabular}
\end{center}
\caption{The values of the parameters of the Glauber models used in this work.}
\label{tab.params}\end{table}
We notice that in both MCGM as well as shMCGM, $n_0$ nearly doubles from $\sNN=200$ GeV to $2.76$ TeV. On the 
other hand, $f$ drops by $20\%$ as we go from RHIC to LHC in MCGM while it stays almost constant in shMCGM. 
However, the shadow parameter $\lambda$ drops by $25\%$ at LHC compared to highest RHIC energy. This suggests 
that $\lambda$ increases with decrease in $\sNN$ and hence the distinction between the two Glauber approaches 
will become even more significant at lower energies (e.g. GSI-FAIR energies).

\begin{figure}[]
\begin{center}
\includegraphics[angle=-90, scale=0.32]{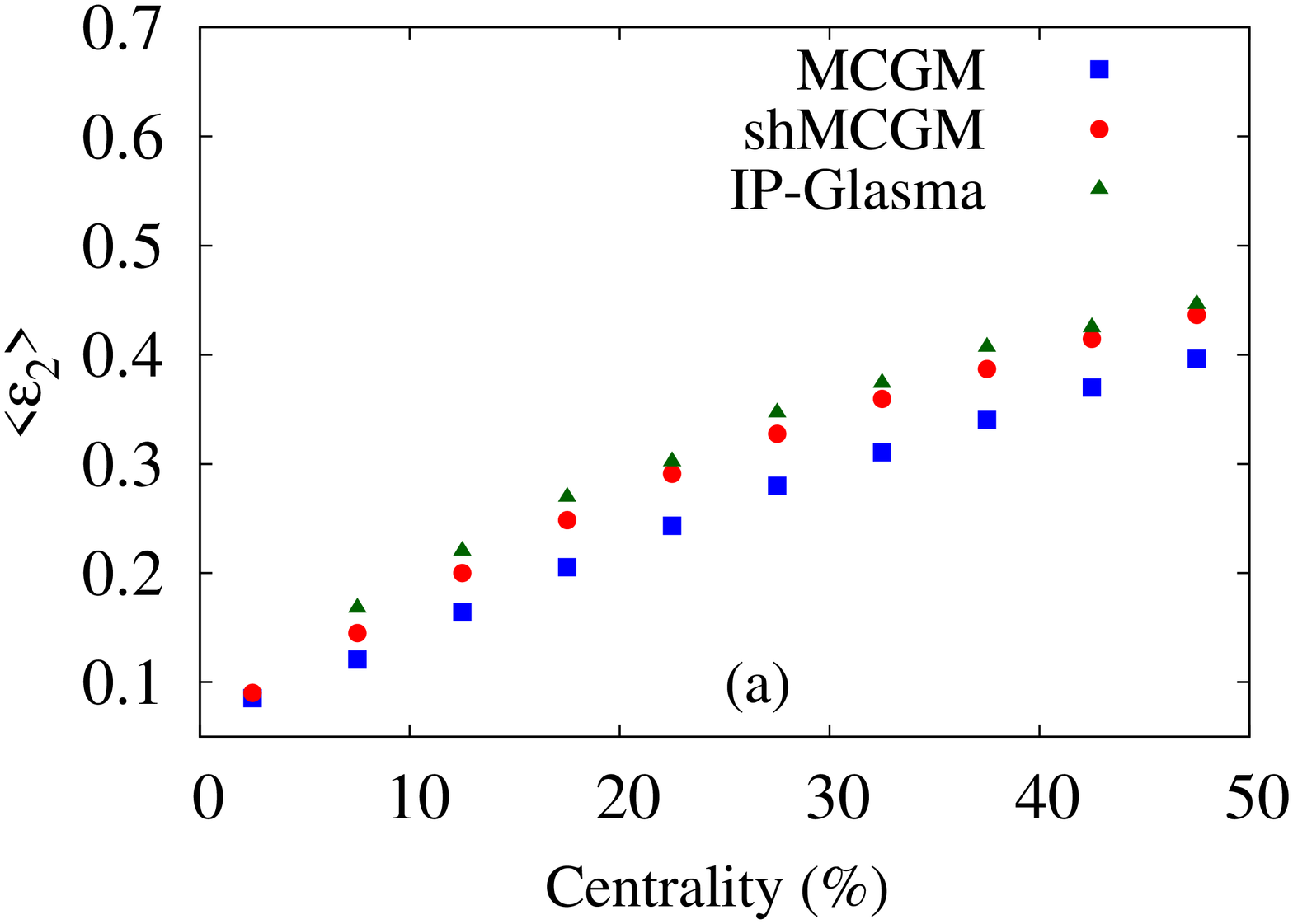} \\
\includegraphics[angle=-90, scale=0.32]{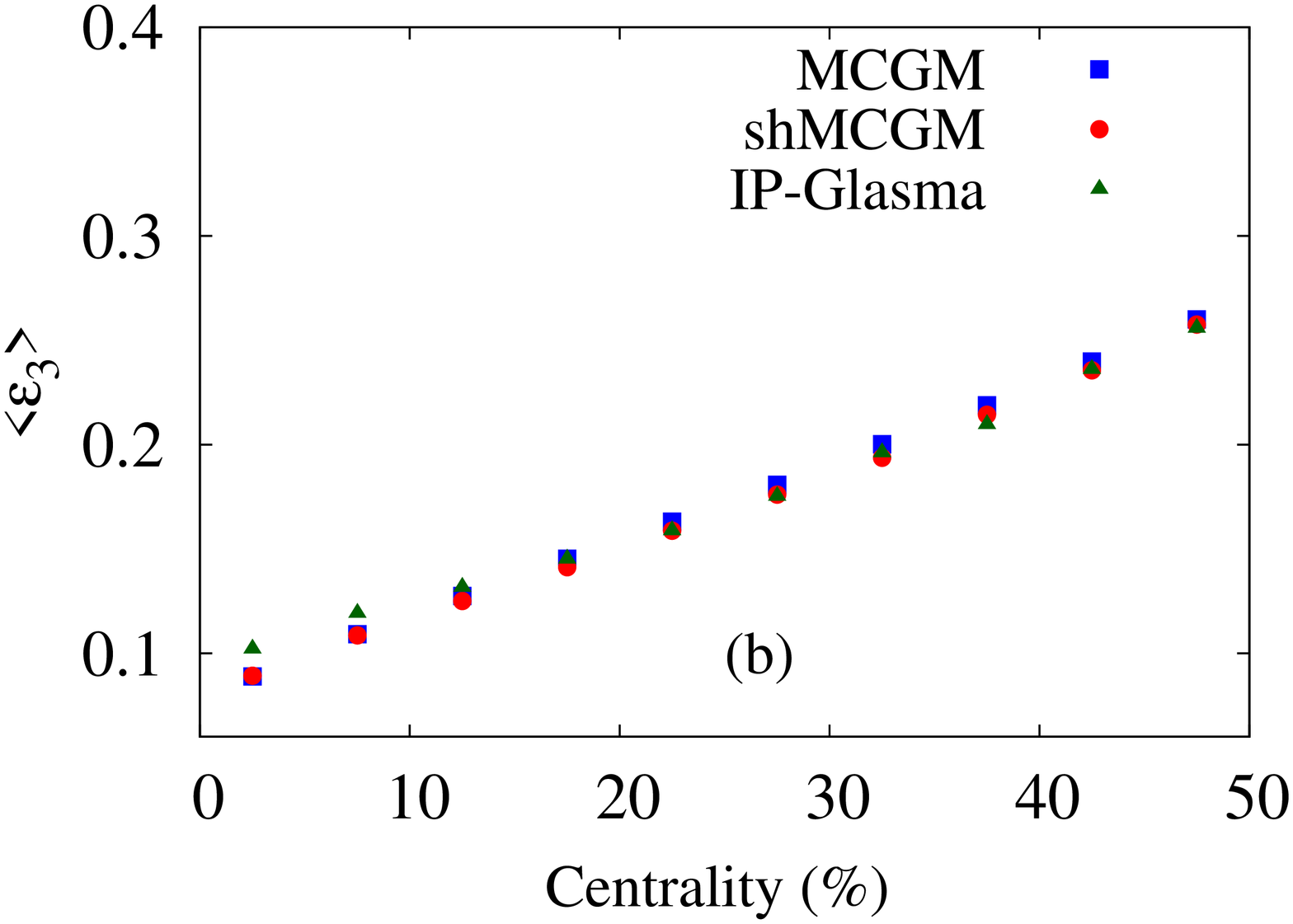} \\
\includegraphics[angle=-90, scale=0.32]{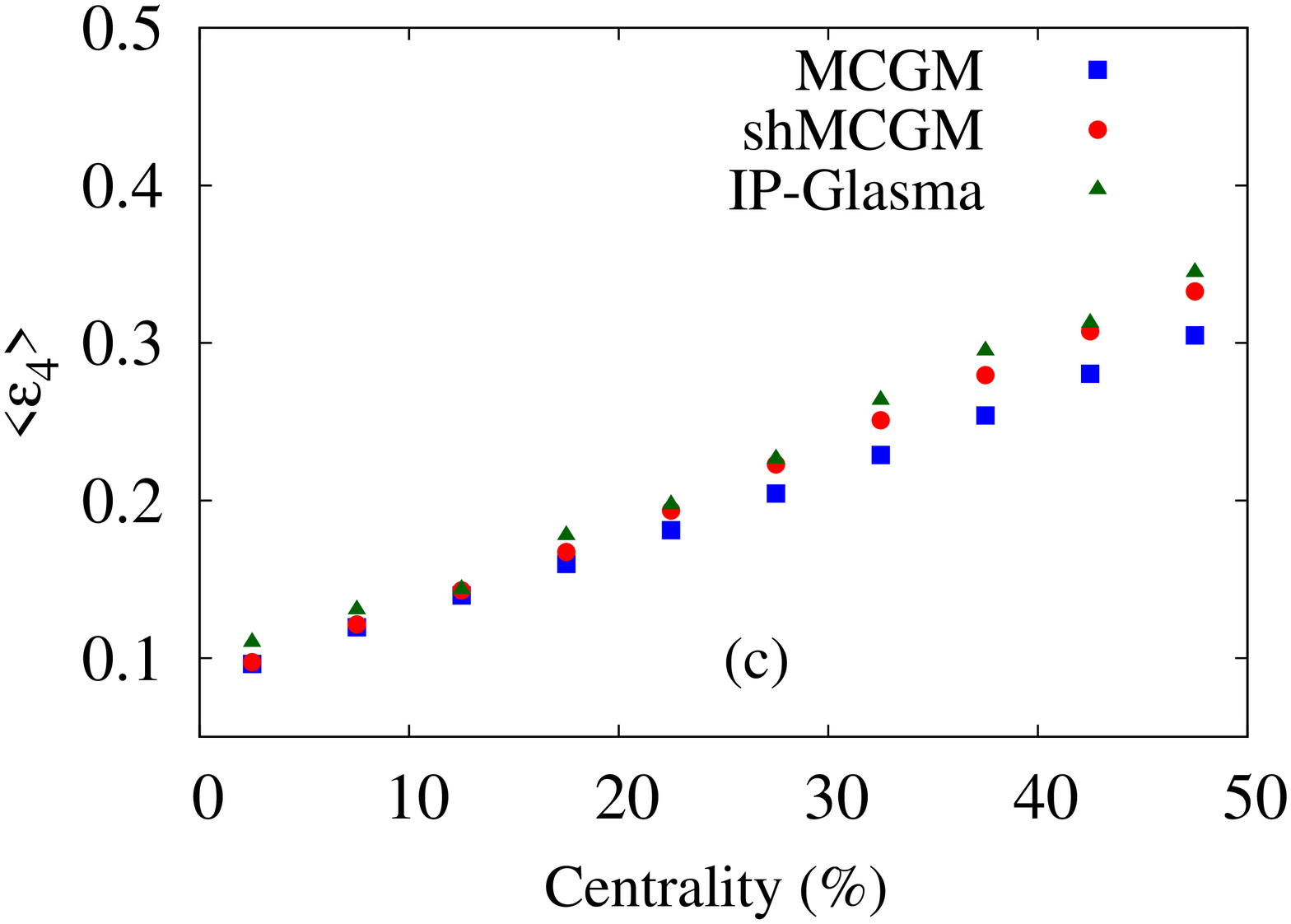}
\end{center}
\caption{(Color online) The centrality dependence of eccentricities compared between IP-Glasma, MCGM and shMCGM.}
\label{fig.meanen}\end{figure}
Thus, having fixed the parameters of the Glauber models, we now look at the model predictions for other 
observables and compare them with data as well as results from other dynamical models of IC like IP-Glasma. We first study 
the centrality dependence of the mean eccentricity harmonics $\varepsilon_n$ of the initial energy deposited 
in the overlap region
\beqa
\varepsilon_ne^{i\Psi_n} &=& \frac{\la r^ne^{in\phi}\ra}{\la r^n\ra}
\label{eq.ecc}
\eeqa
where $n=2,3,..$. The coordinate system is chosen such that $\la \vec{r}\ra=0$. $\la ... \ra$ represents 
averaging over the transverse plane with the initial energy deposited on the transverse plane $\epsilon\l x,y\r$ 
as the weight function. In Fig.~\ref{fig.meanen}, we have plotted the ensemble average of 
$\varepsilon_2$, $\varepsilon_3$ and $\varepsilon_4$ vs their centrality which is determined from the 
final state (FS) charged hadron multiplicity. We have shown the results for MCGM, shMCGM and IP-Glasma. All the 
models show a rising trend for $\varepsilon_2$ with centrality which is also expected from geometrical 
arguments- events with lower multiplicity occur with larger impact parameter which results in larger ellipticity 
of the overlap region. For all centralities, $\varepsilon_2$ in shMCGM is enhanced as compared to MCGM and is 
also in good agreement with IP-Glasma results. The higher value of $\varepsilon_2$ in shMCGM as compared to 
MCGM was also found at RHIC energy and is a typical effect due to nucleon shadowing~\cite{Chatterjee:2015aja}. 
The ends of the minor axis of the collision zone have on an average larger number of participants as compared to 
the ends of the major axis of the overlap region which calls for larger shadowing effect at the ends of the 
minor axis. This effectively reduces the minor axis more than the major axis, thus increasing the $\varepsilon_2$ 
of the overlap region.

The higher harmonics arise due to granularity in the IC which is controlled by $N_{part}$: smaller 
$N_{part}$ corresponds to larger granularity~\cite{Bhalerao:2011bp,Schenke:2014tga}. This explains the 
rising trend in $\varepsilon_3$ and $\varepsilon_4$ with centrality. We note that both MCGM and shMCGM 
yield almost similar values for these higher eccentricities even though in shMCGM the effective number 
of participants is smaller than in MCGM due to the shadowing effect. This is so because the shadowing effect 
systematically weakens only those sources which have other sources in front. Thus, weakening of these 
sources in the bulk do not increase the granularity in the transverse plane.

\begin{figure}[]
\begin{center}
\includegraphics[angle=-90, scale=0.32]{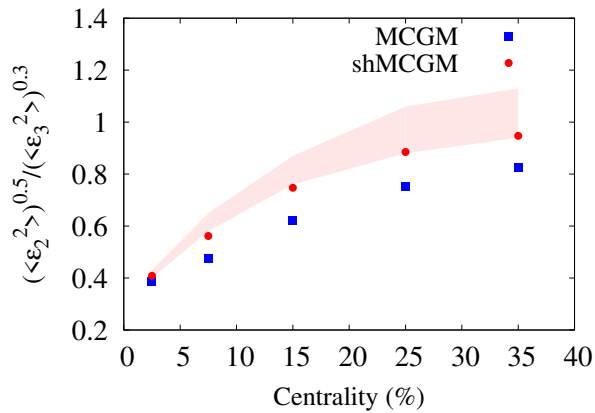}
\end{center}
\caption{(Color online) The centrality dependence of the ratio of r.m.s. $\varepsilon_2$ to 
$\varepsilon_3$ compared between MCGM and shMCGM. Also shown is the band as proposed in 
Ref.~\cite{Retinskaya:2013gca} that is required to explain the correlation of $v_2-v_3$ in data assuming 
linear hydrodynamic response.}
\label{fig.e2e3}\end{figure}
The correlation between the even-odd harmonics largely stem from the $\varepsilon_2-\varepsilon_3$ 
correlation of the initial state (IS). Starting from the observed correlation in the data of $v_2-v_3$ 
at the LHC, an allowed band for the ratio of r.m.s values of $\varepsilon_2$ to $\varepsilon_3$ was 
obtained in Ref.~\cite{Retinskaya:2013gca} within the realm of linear response. In Fig.~\ref{fig.e2e3} 
we have shown this band. We also show the values obtained for the same quantity in MCGM and shMCGM. 
The enhancement of $\varepsilon_2$ in shMCGM as compared to MCGM as noted earlier in Fig.~\ref{fig.meanen} 
also helps here- it pushes the prediction for the ratio of r.m.s. of $\varepsilon_2$ to $\varepsilon_3$ 
into the band that is favored by data unlike the case of MCGM which underpredicts as compared to the band.

\begin{figure}[]
\begin{center}
\includegraphics[angle=-90, scale=0.32]{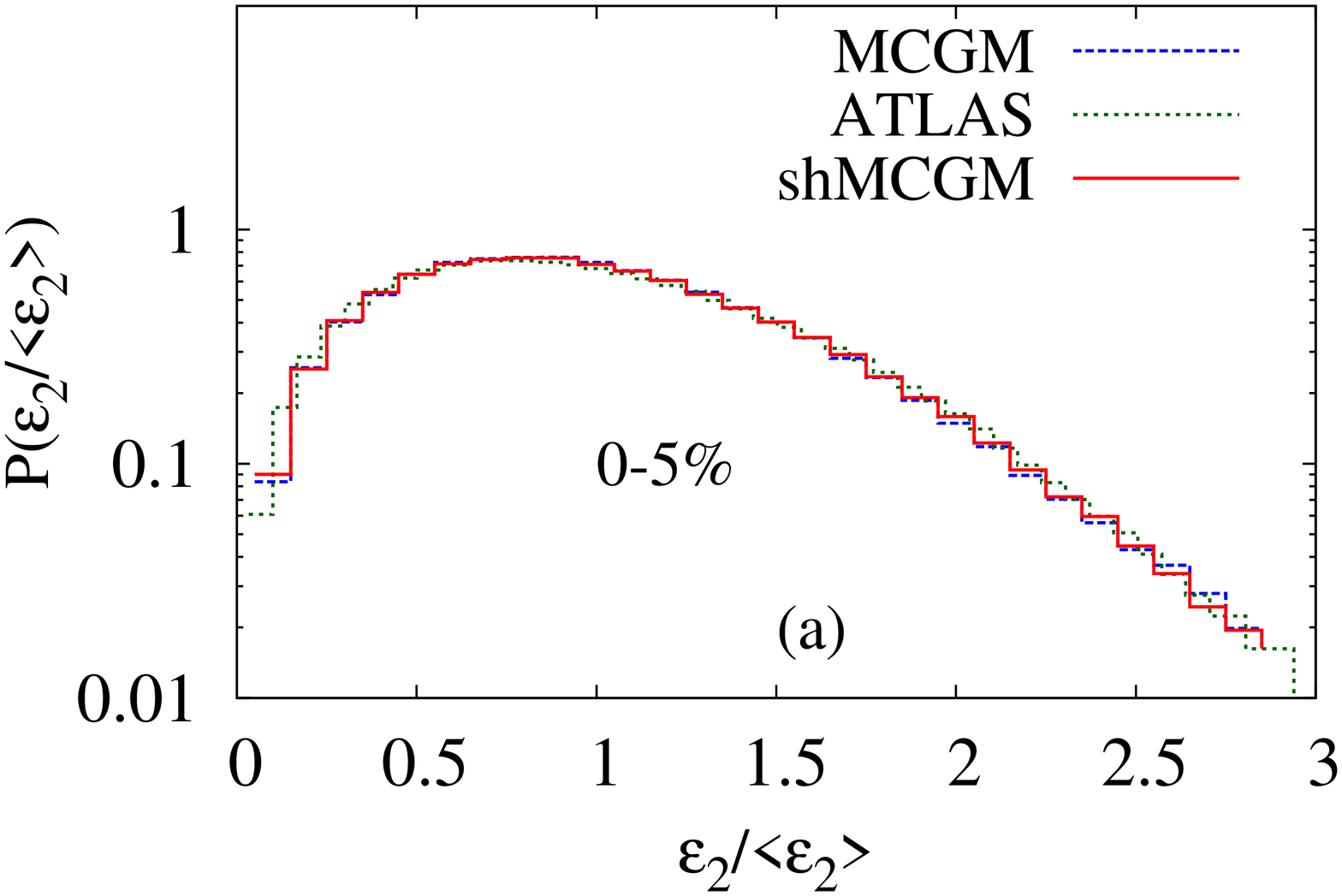} \\
\includegraphics[angle=-90, scale=0.32]{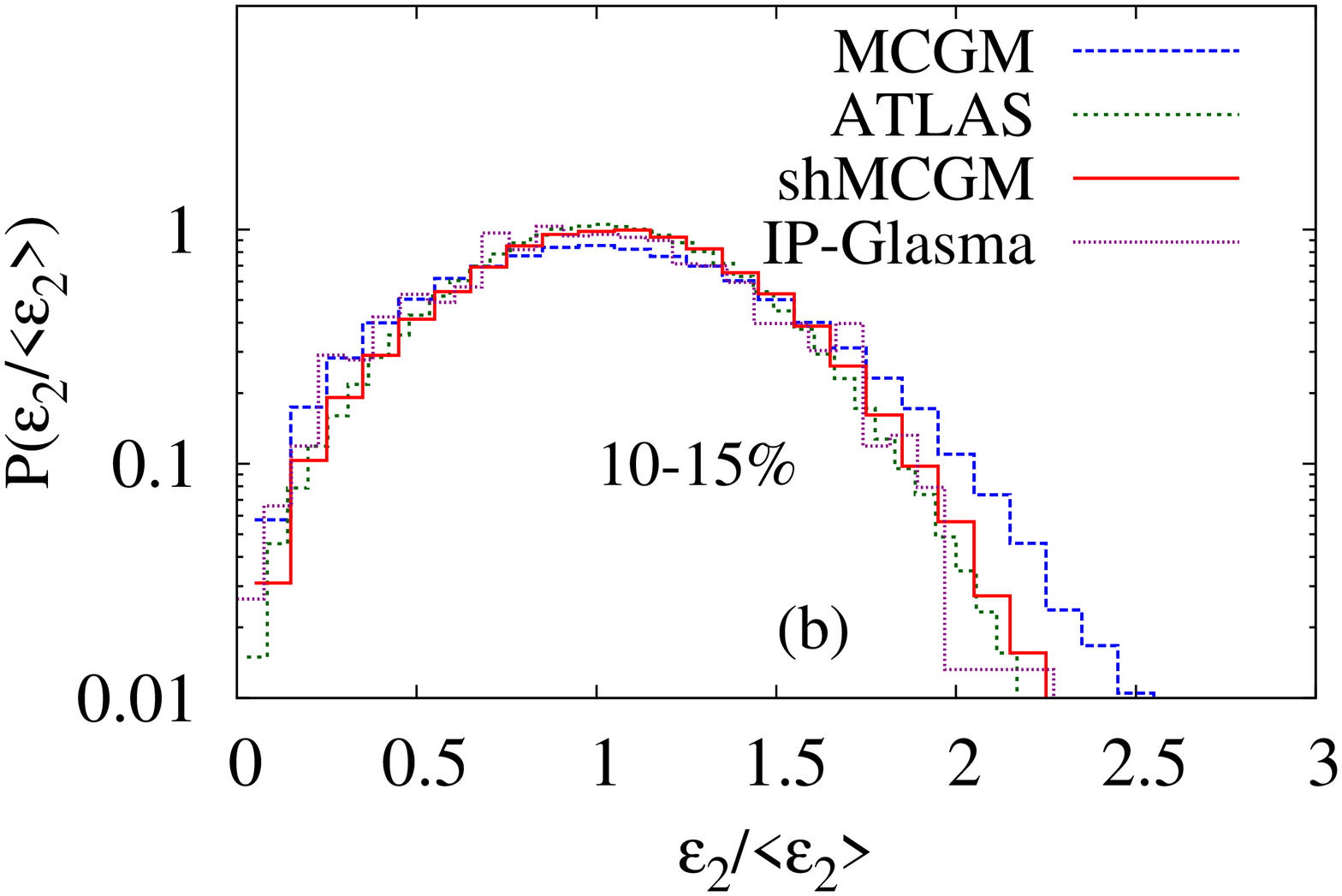} \\
\includegraphics[angle=-90, scale=0.32]{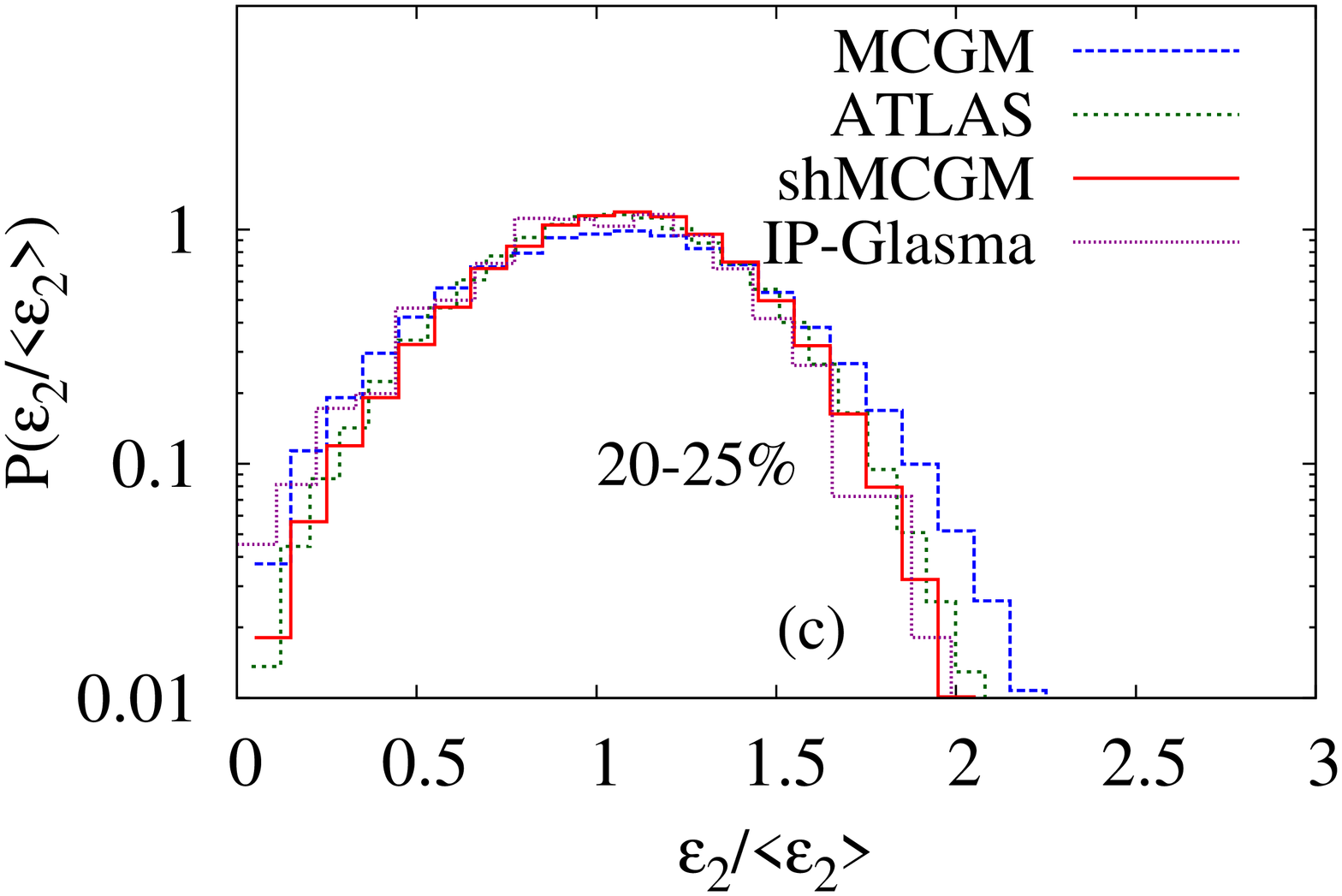}
\end{center}
\caption{(Color online) The E/E distribution of $\varepsilon_2$ compared 
between data~\cite{Timmins:2013hq,Aad:2013xma}, IP-Glasma, MCGM and shMCGM.}
\label{fig.ebyee2}\end{figure}

\begin{figure}[]
\begin{center}
\includegraphics[angle=-90, scale=0.32]{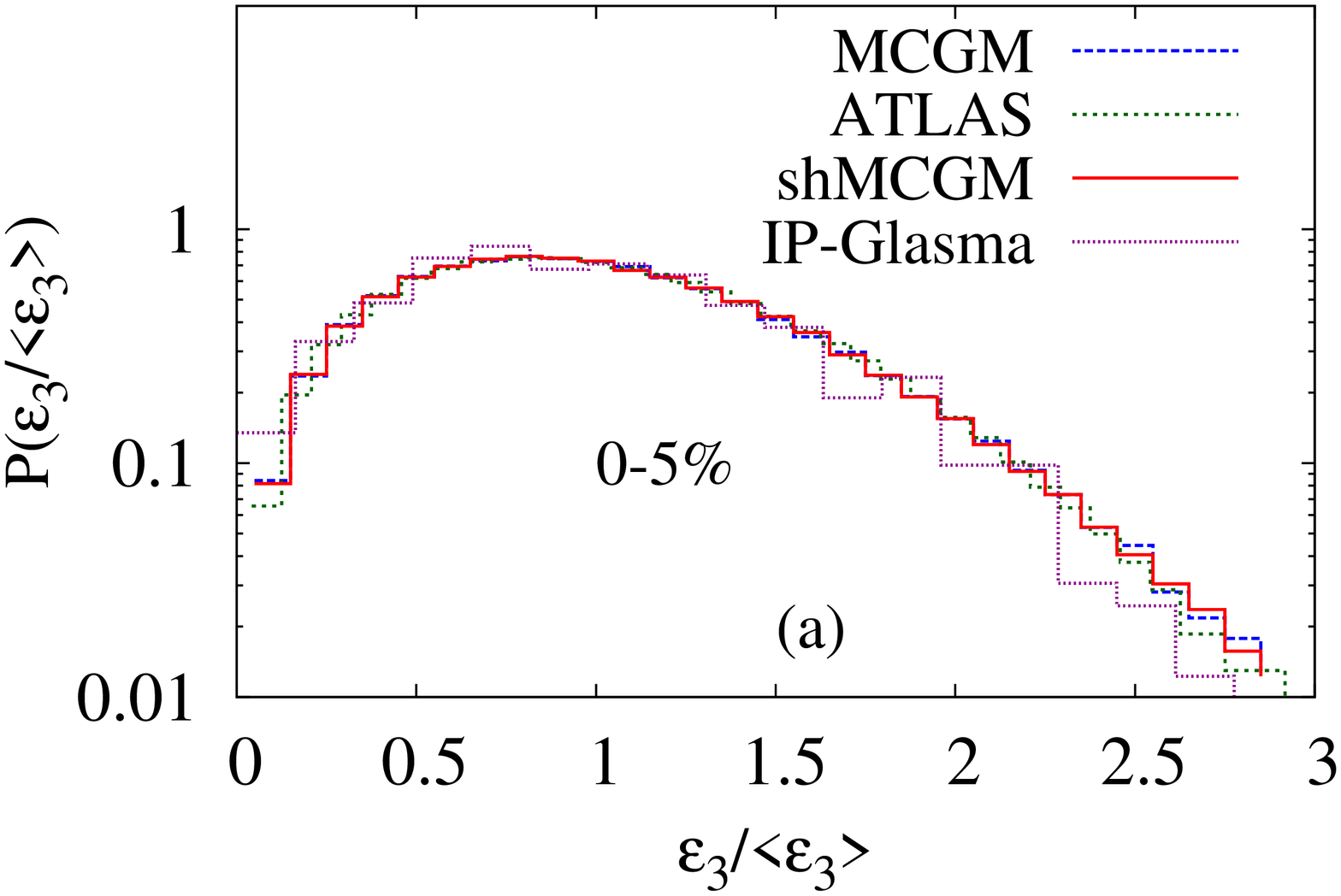} \\
\includegraphics[angle=-90, scale=0.32]{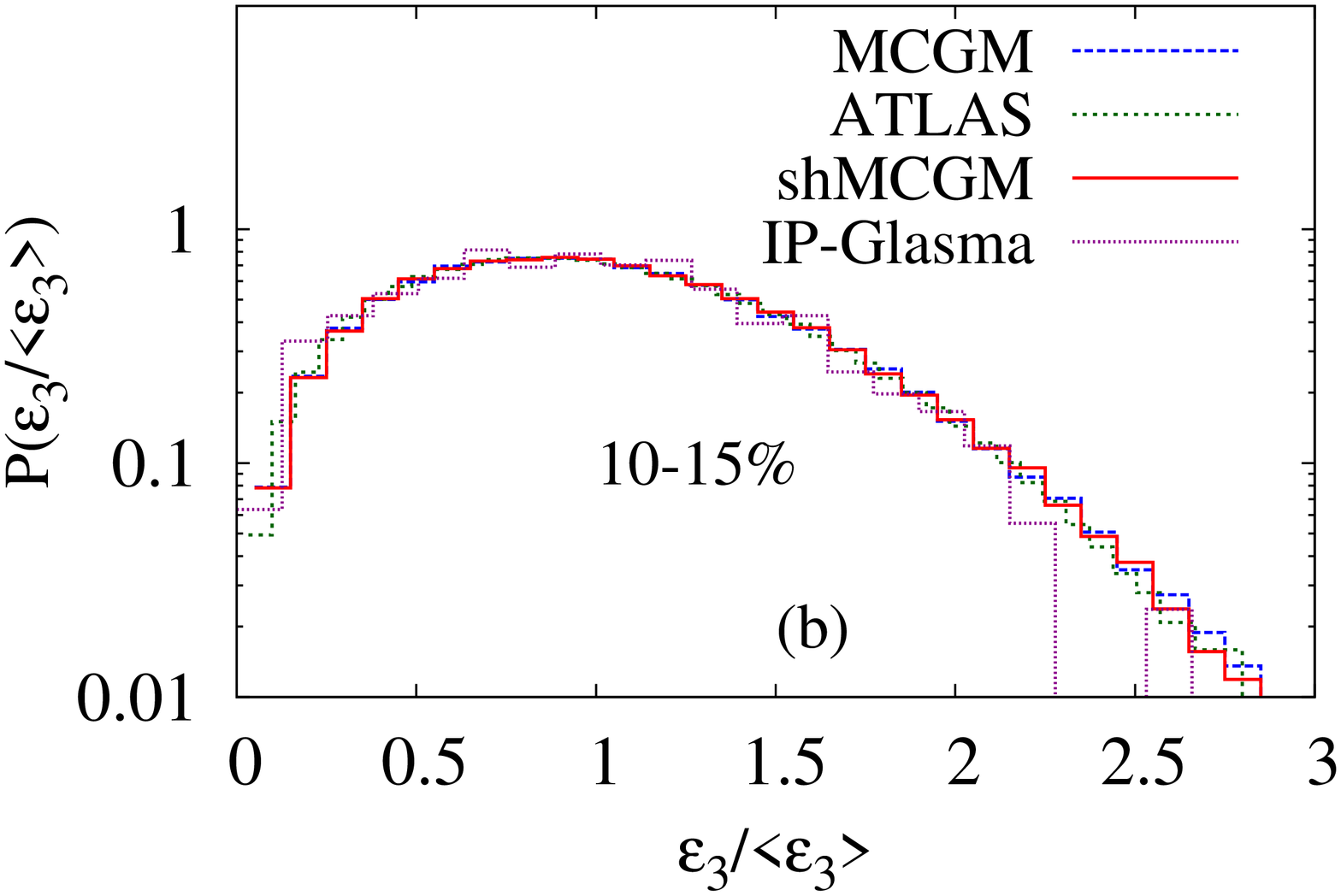} \\
\includegraphics[angle=-90, scale=0.32]{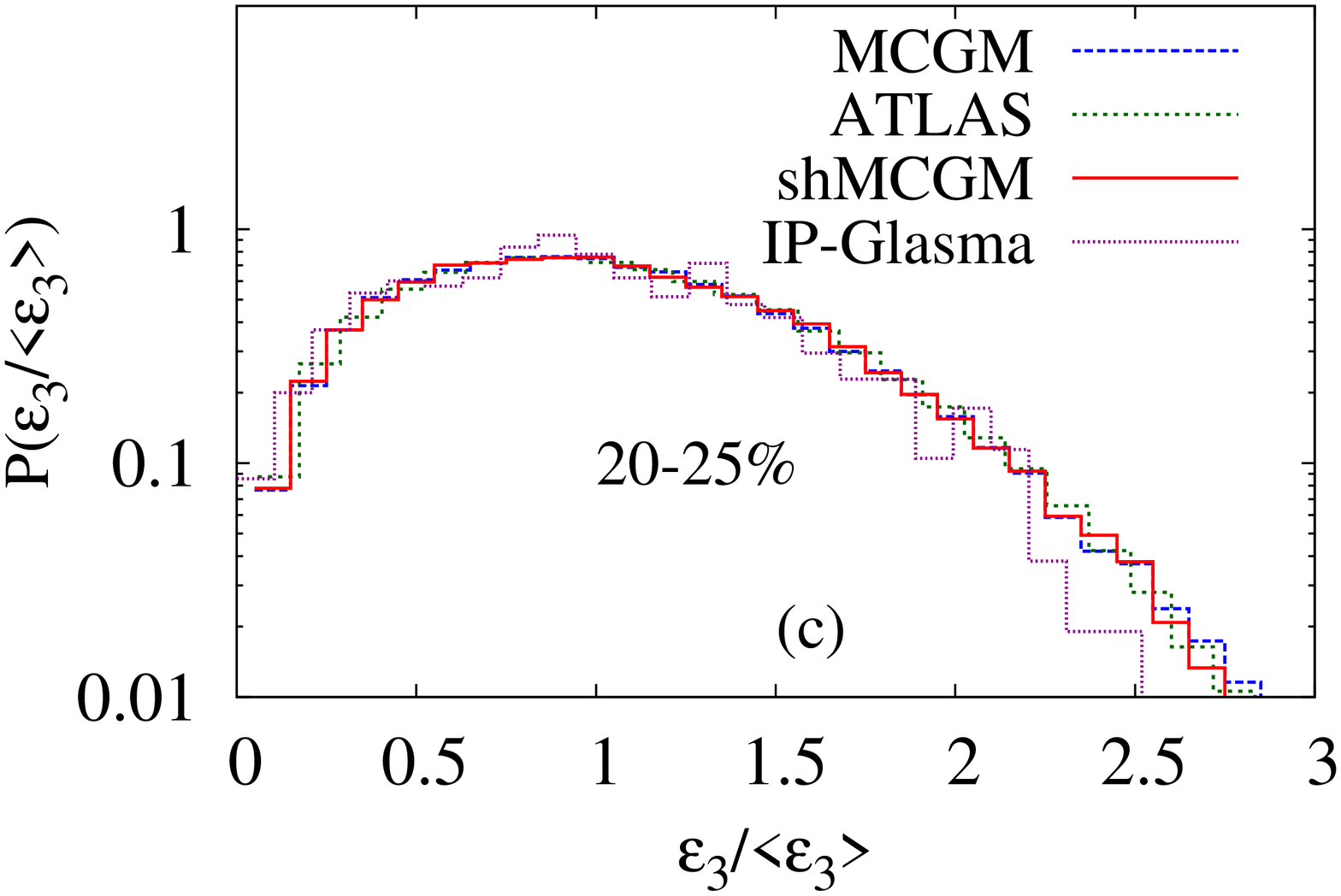}
\end{center}
\caption{(Color online) The E/E distribution of $\varepsilon_3$ compared 
between data~\cite{Timmins:2013hq,Aad:2013xma}, IP-Glasma, MCGM and shMCGM.}
\label{fig.ebyee3}\end{figure}

\begin{figure}[]
\begin{center}
\includegraphics[angle=-90, scale=0.32]{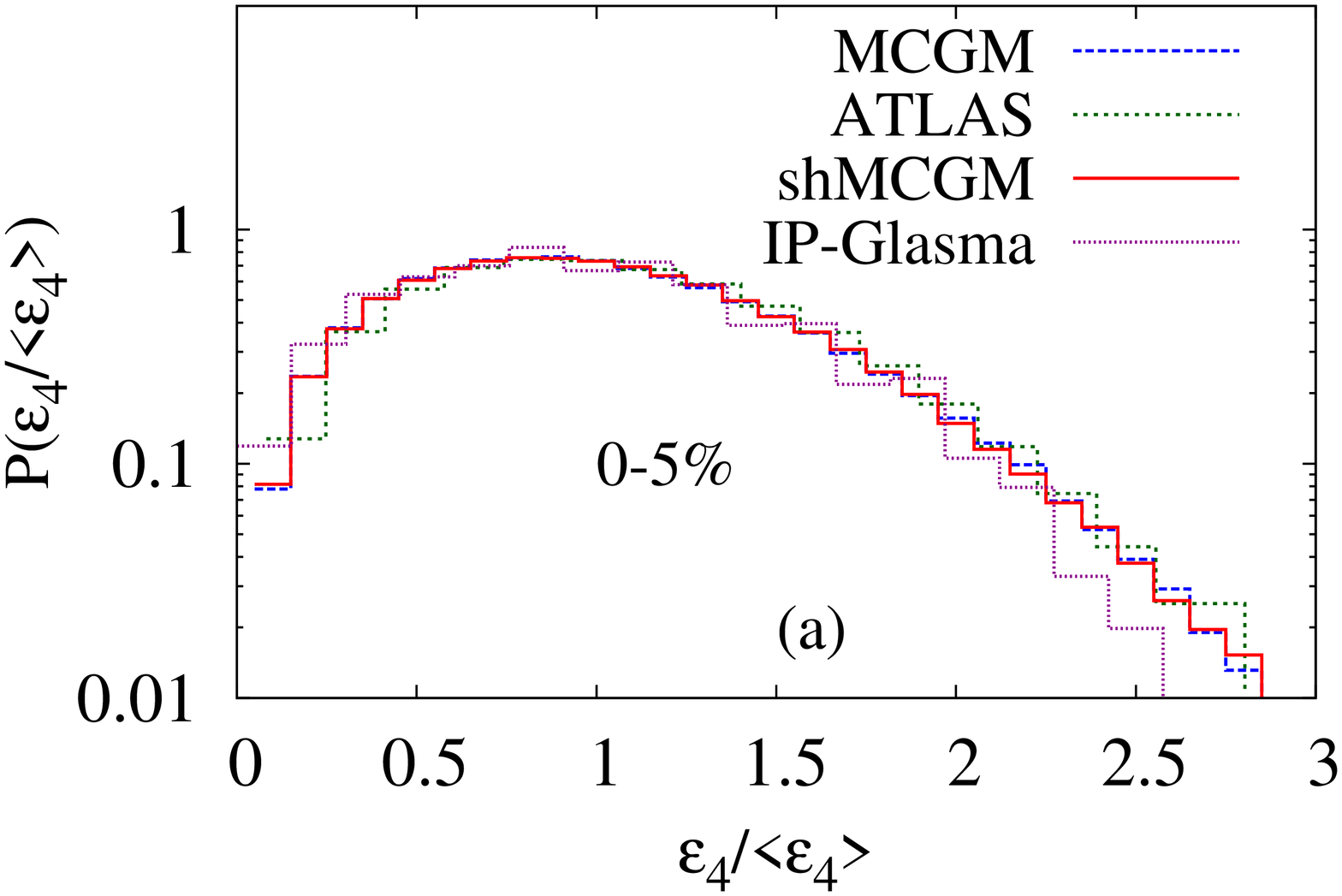} \\
\includegraphics[angle=-90, scale=0.32]{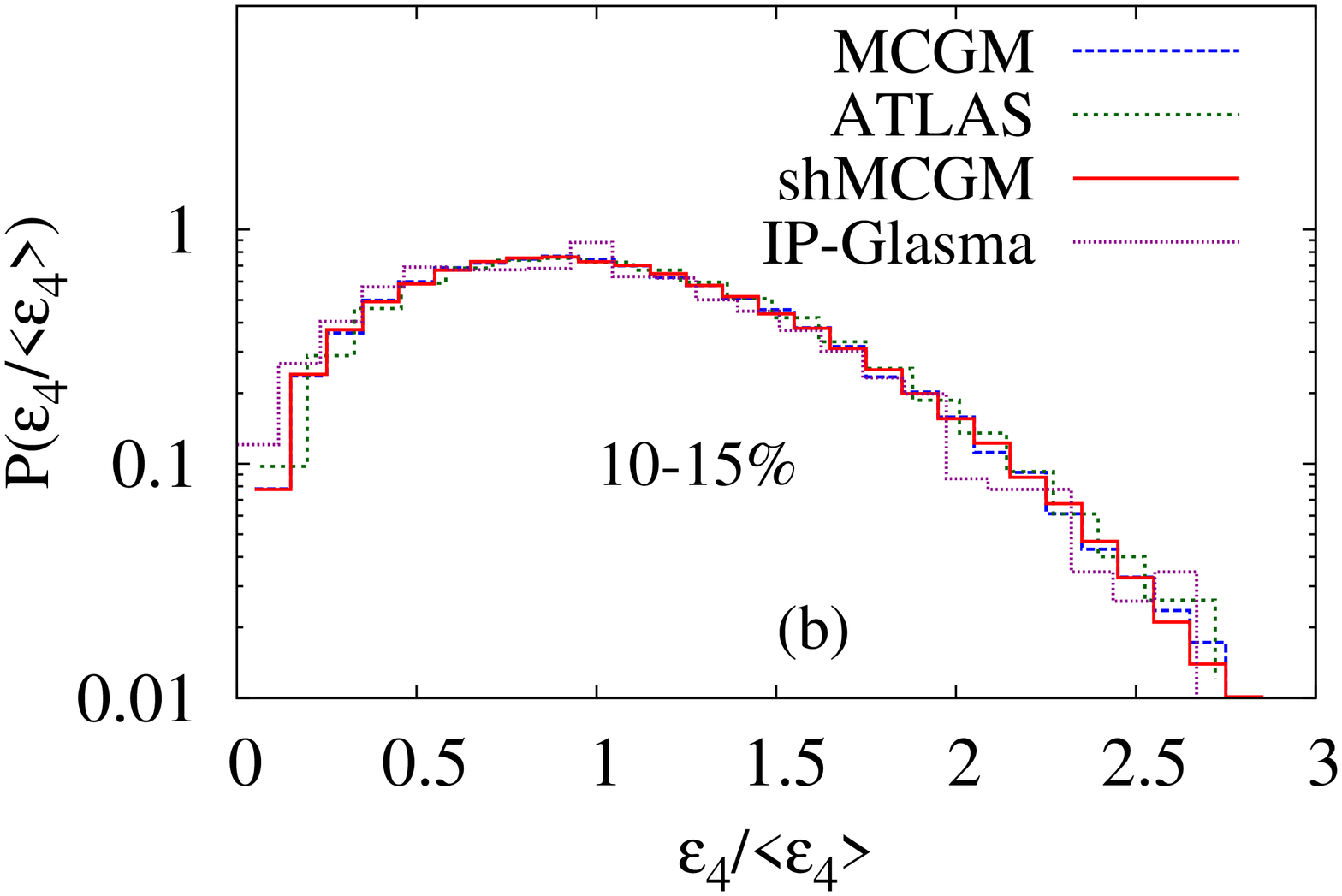} \\
\includegraphics[angle=-90, scale=0.32]{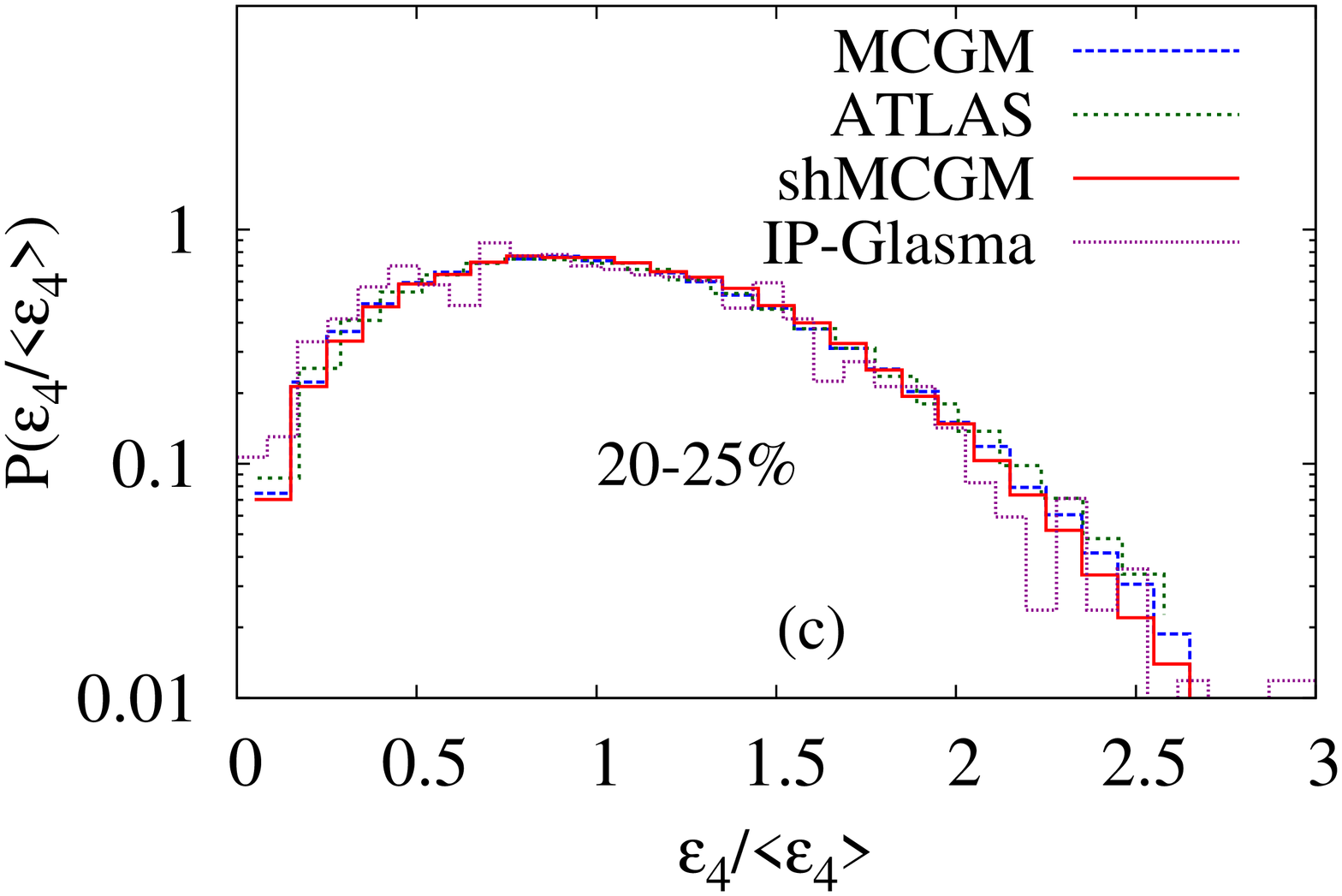}
\end{center}
\caption{(Color online) The E/E distribution of $\varepsilon_4$ compared 
between data~\cite{Timmins:2013hq,Aad:2013xma}, IP-Glasma, MCGM and shMCGM.}
\label{fig.ebyee4}\end{figure}
We now turn our attention from the mean geometric properties in the IC to their fluctuations. We first 
analyse the E/E distributions of the $\varepsilon_n$ scaled by their ensemble average values and compare 
with that of IP-Glasma~\cite{Gale:2012rq} as well as ATLAS data of $v_n$~\cite{Timmins:2013hq,Aad:2013xma}. 
As long as the hydrodynamic response is linear ($v_n = k_n\epsilon_n$ where $k_n$ is a constant), we 
expect the E/E distributions of $\varepsilon_n/\la\epsilon_n\ra$ to be a good representative of $v_n/\la v_n\ra$. 
In Figs.~\ref{fig.ebyee2}, \ref{fig.ebyee3} and \ref{fig.ebyee4} we have plotted the E/E distribution plots 
for $\varepsilon_2$, $\varepsilon_3$ and $\varepsilon_4$ for the following centrality classes: $\l0-5\r\%$, 
$\l10-15\r\%$ and $\l20-25\r\%$. Overall, there is good quantitative agreement between shMCGM and data as 
well as IP-Glasma. It is well known that the standard MCGM produces a broader E/E distribution as compared 
to data as well as IP-Glasma results~\cite{Schenke:2014tga, Niemi:2015qia}. However as already argued earlier 
in Ref.~\cite{Chatterjee:2015aja}, the shadowing effect {\it{shadows}} the participants as well as their E/E 
fluctuations in position which eventually results in narrower E/E distribution that is in good agreement with 
data and IP-Glasma predictions.

\begin{figure}[]
\begin{center}
\includegraphics[angle=-90, scale=0.32]{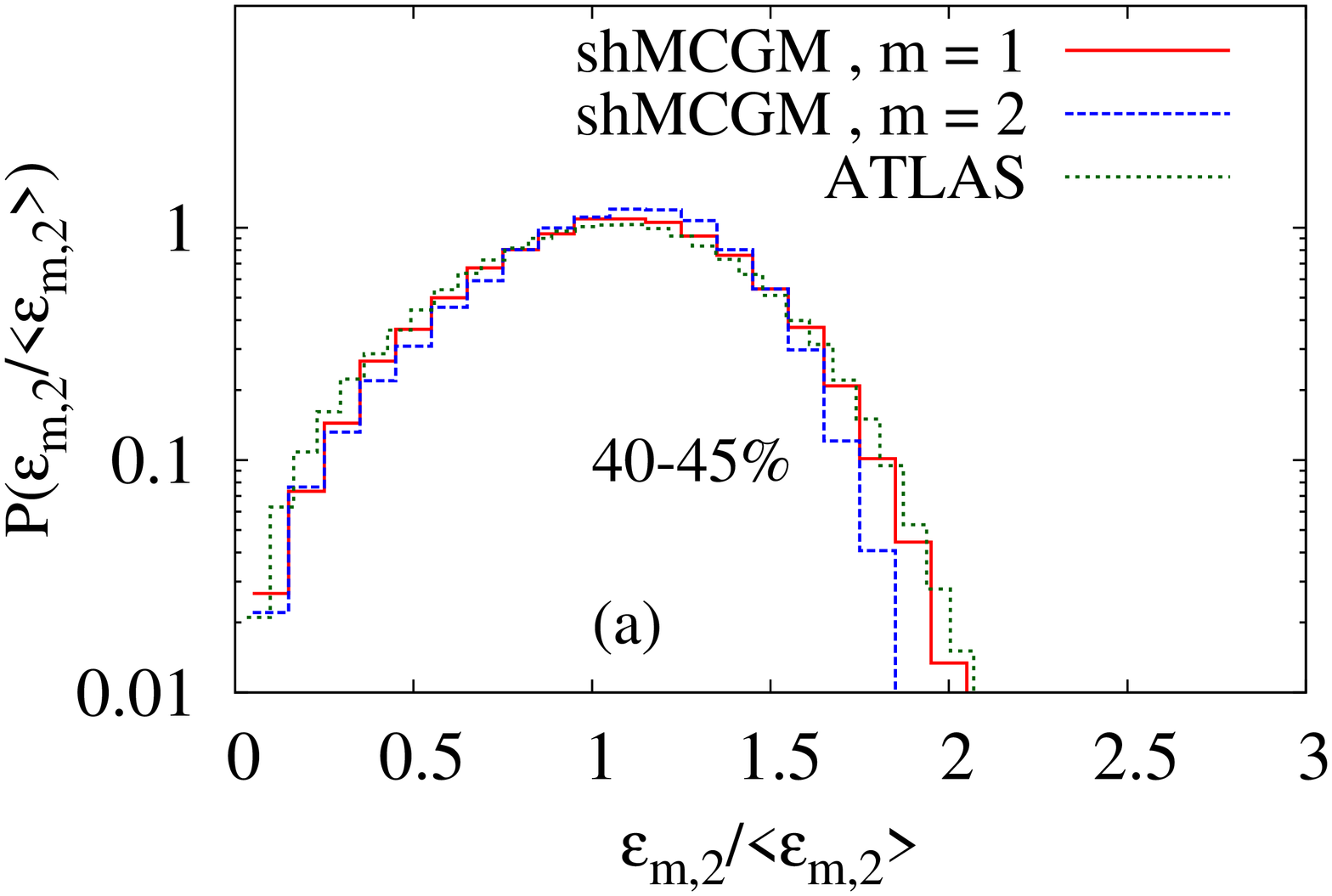} \\
\includegraphics[angle=-90, scale=0.32]{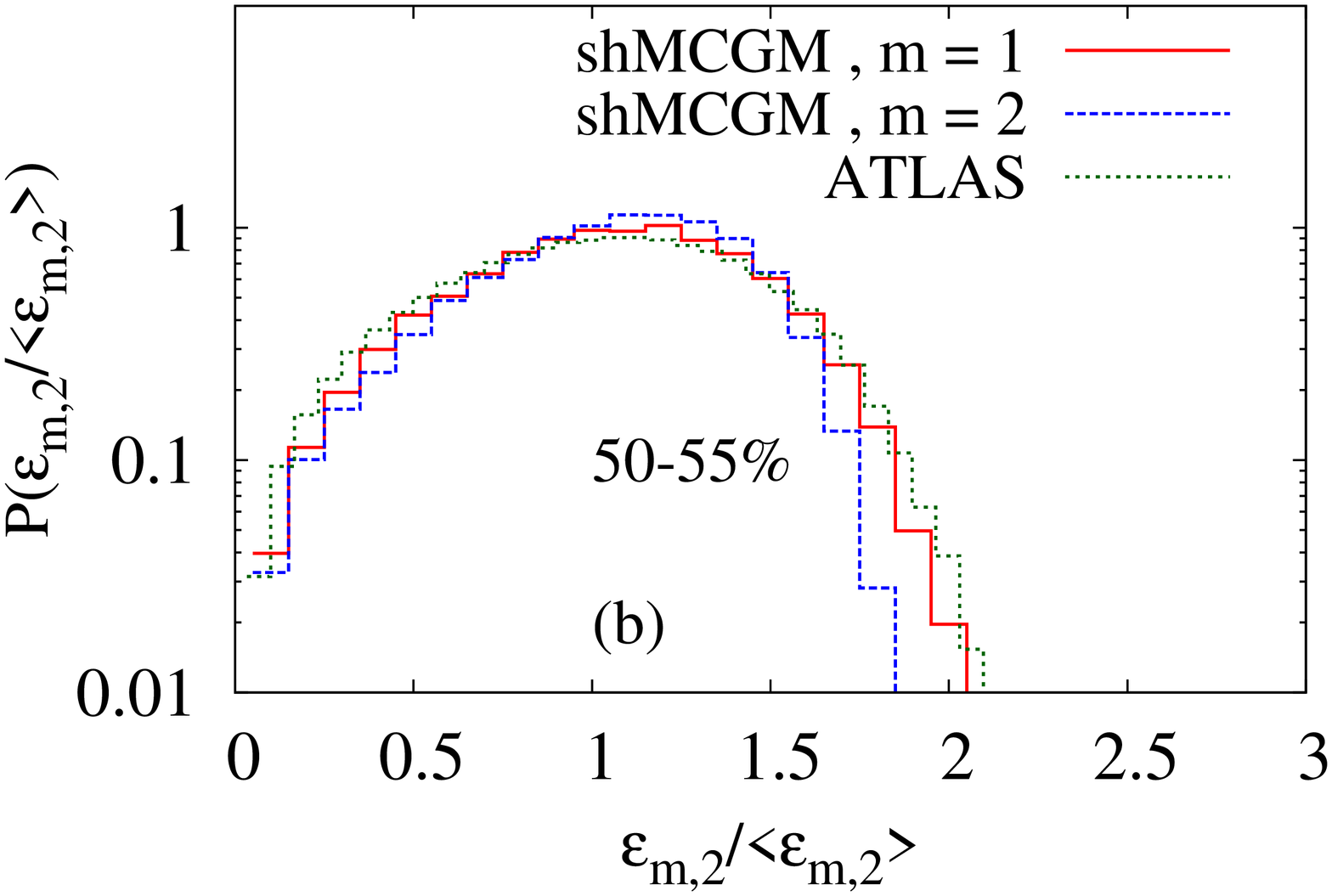}
\end{center}
\caption{(Color online) The comparison between E/E distribution of $\varepsilon_{1,2}$ and 
$\varepsilon_2$ obtained in shMCGM with data of $v_2$~\cite{Timmins:2013hq,Aad:2013xma}.}
\label{fig.ebyee12}\end{figure}
For further peripheral centralities, the agreement in case of the E/E distributions of $\varepsilon_2$ worsen. 
On the other hand, the distributions of the higher harmonics continue to be in good agreement. Recently, this 
has been shown as evidence of cubic hydrodynamic response to ellipticity in peripheral 
collisions~\cite{Noronha-Hostler:2015dbi}. A modified predictor of $v_2$, in terms of $\varepsilon_{1,2}$ was 
shown to do a much better job in the EKRT model for such peripheral bins~\cite{Niemi:2015qia} where it is defined 
as: 
\beqa
\varepsilon_{1,2}e^{i\Psi_{1,2}} &=& \frac{\la re^{i2\phi}\ra}{\la r\ra}
\label{eq.ecc12}
\eeqa
In Fig.~\ref{fig.ebyee12} we have plotted the E/E distribution for $\varepsilon_{1,2}$ and 
compared with the case of $\varepsilon_2$ as well as $v_2$ from data for the centrality bins of 
$\l40-45\r\%$ and $\l50-55\r\%$. We indeed find better agreement between data of $v_2$ and 
shMCGM prediction for $\varepsilon_{1,2}$ than $\varepsilon_2$.

\begin{figure}[]
\begin{center}
\includegraphics[angle=-90, scale=0.32]{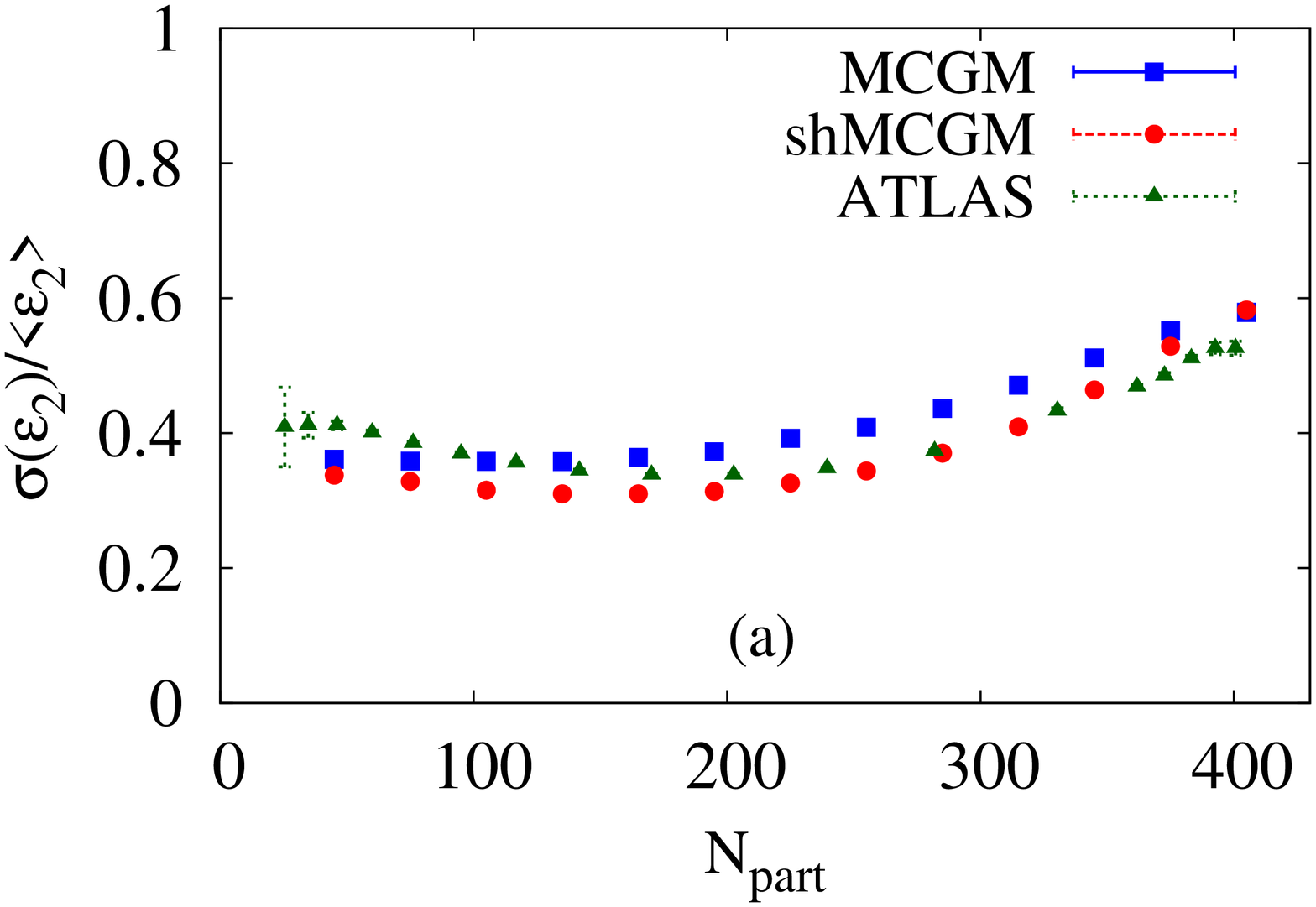} \\
\includegraphics[angle=-90, scale=0.32]{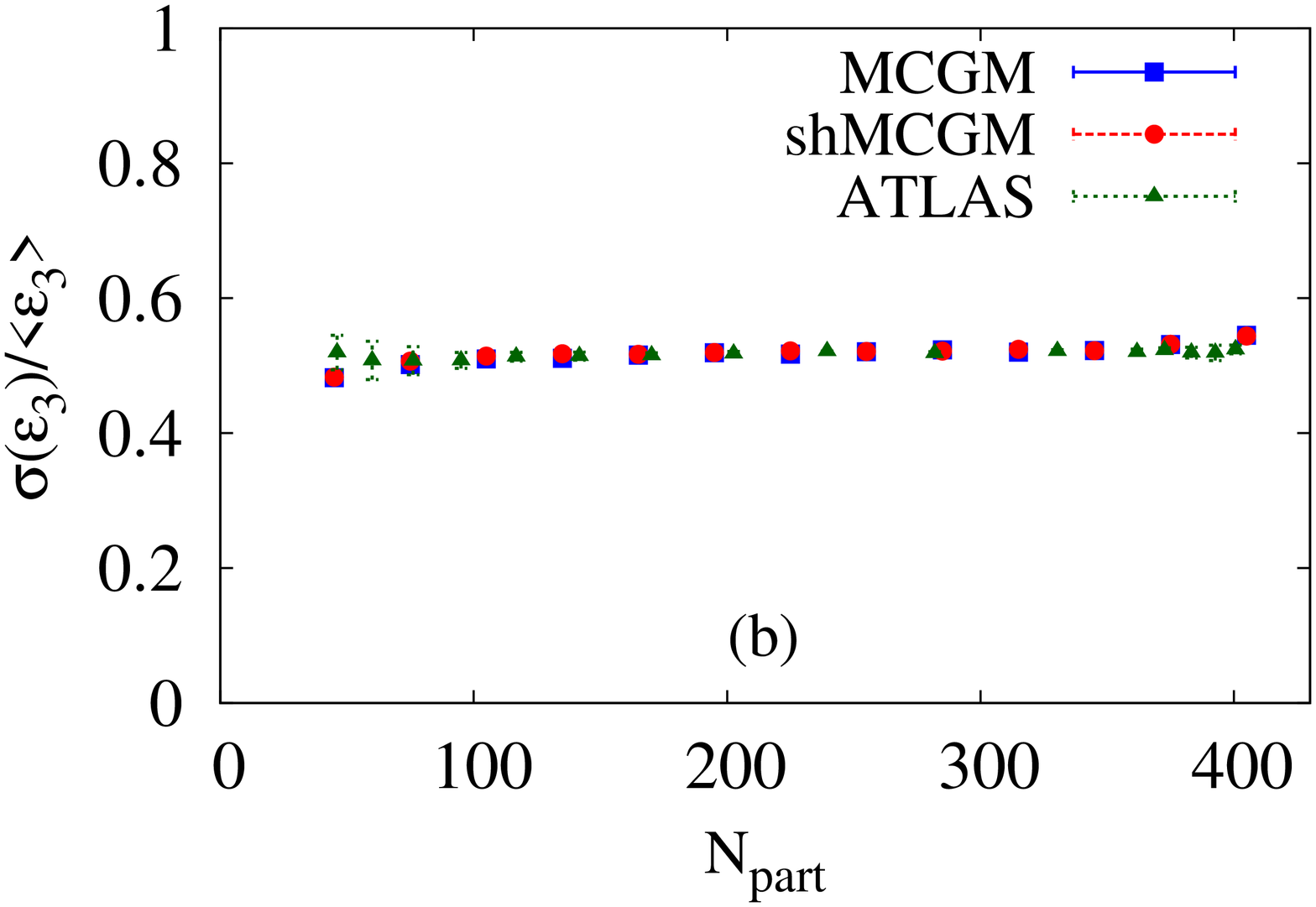} \\
\includegraphics[angle=-90, scale=0.32]{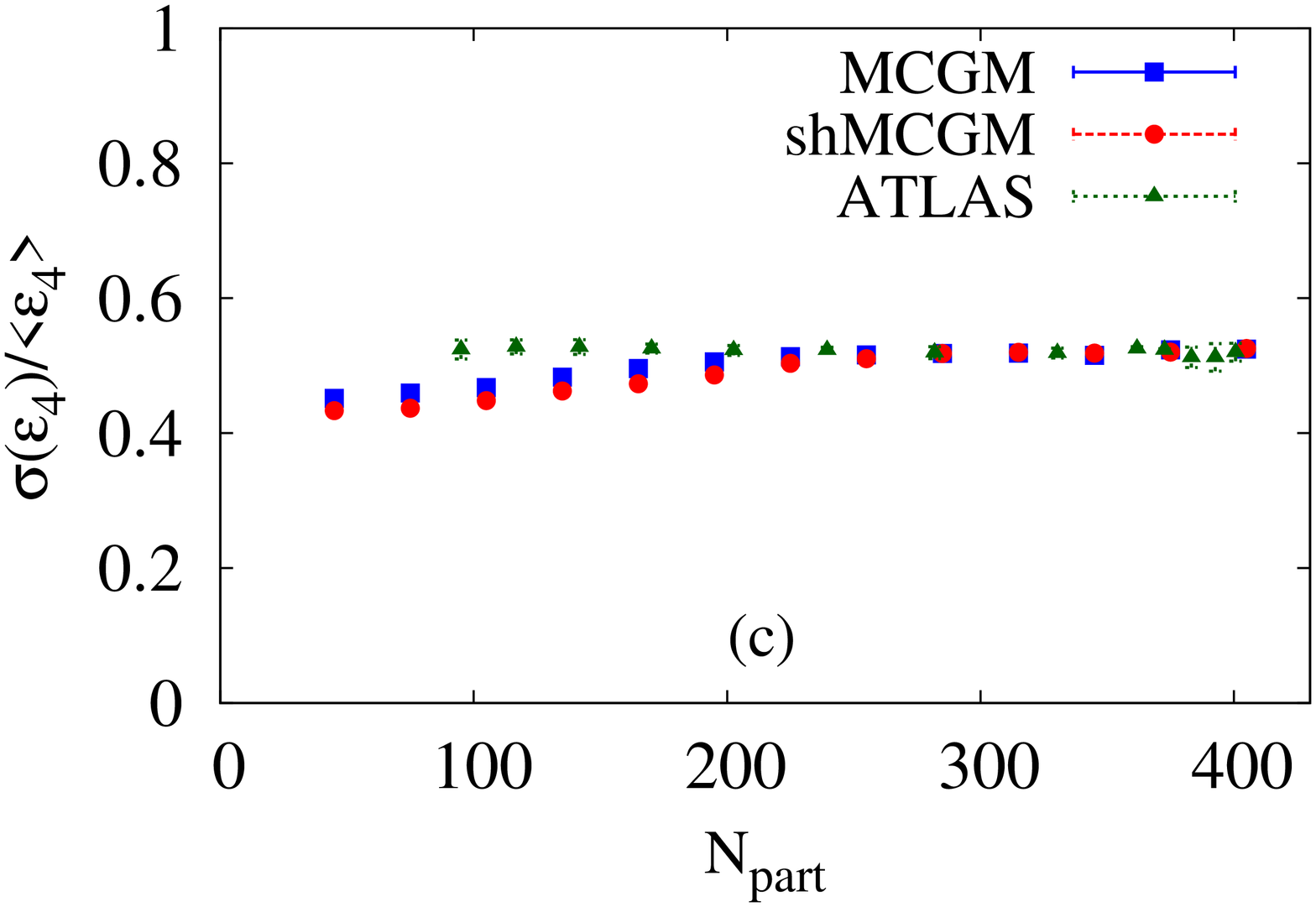} 
\end{center}
\caption{(Color online) The centrality dependence of normalised standard deviation of the eccentricities compared 
between data~\cite{Aad:2013xma}, MCGM and shMCGM.}
\label{fig.sigmaen}\end{figure}

\begin{figure}[]
\begin{center}
\includegraphics[angle=-90, scale=0.32]{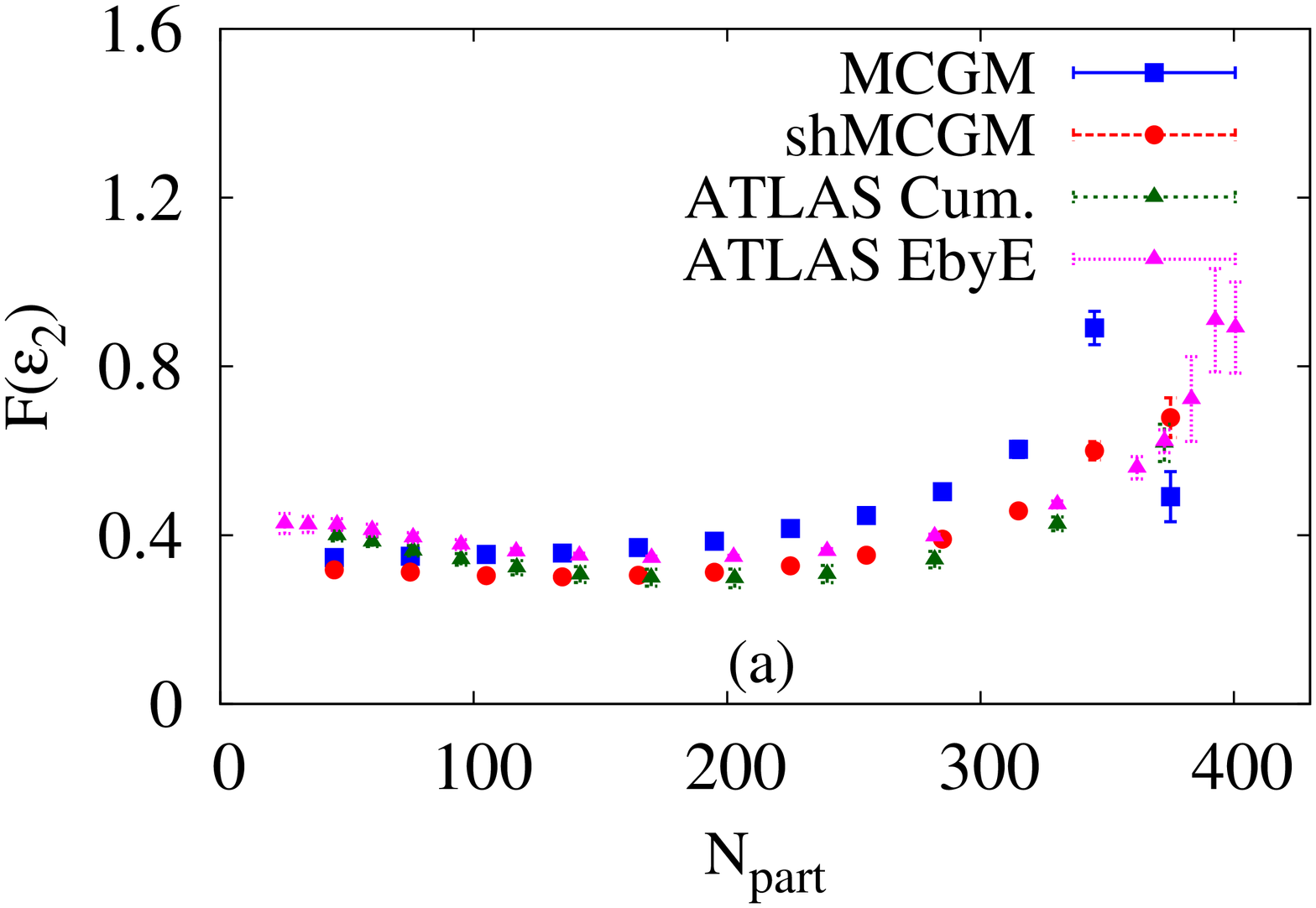} \\
\includegraphics[angle=-90, scale=0.32]{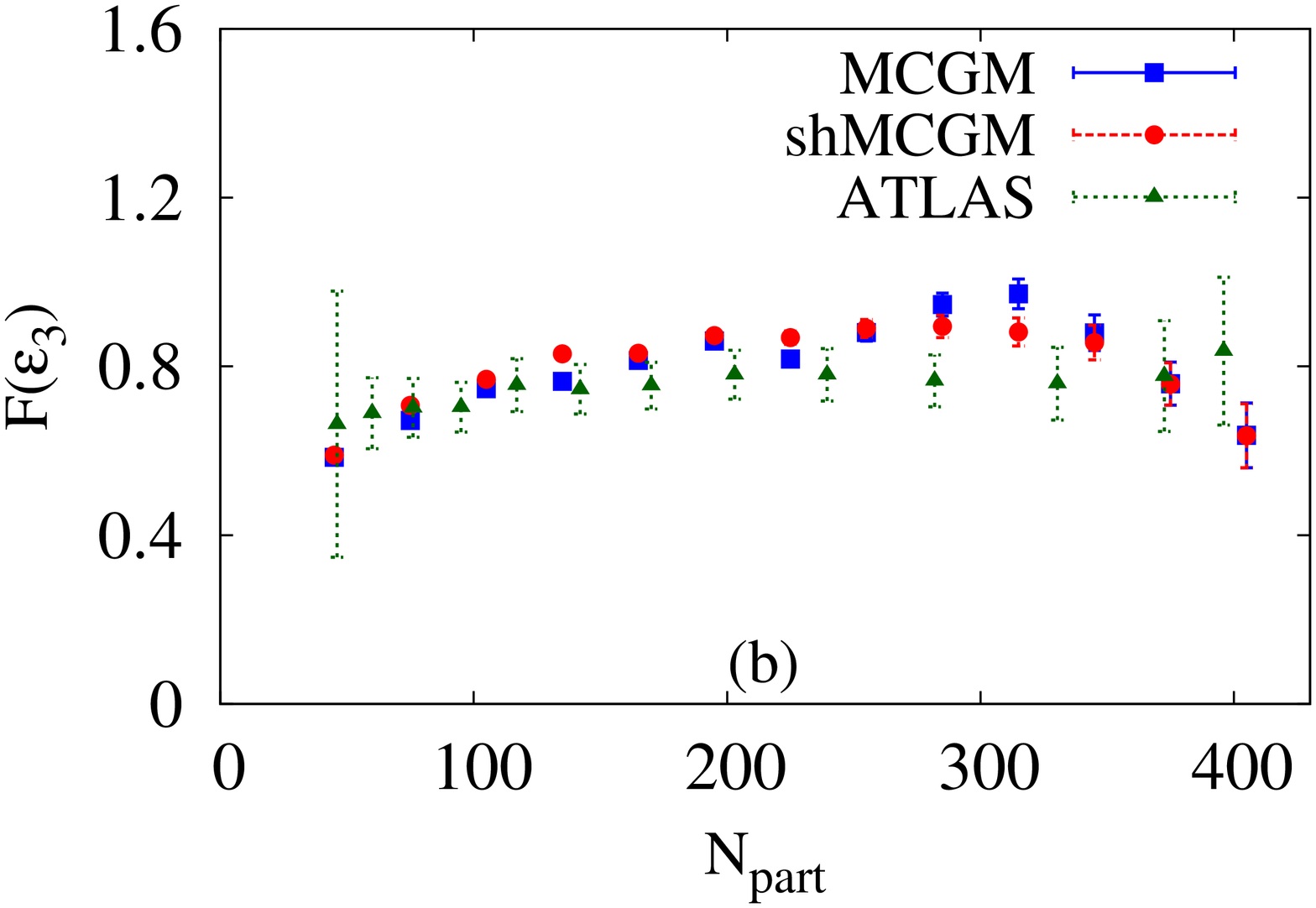} \\
\includegraphics[angle=-90, scale=0.32]{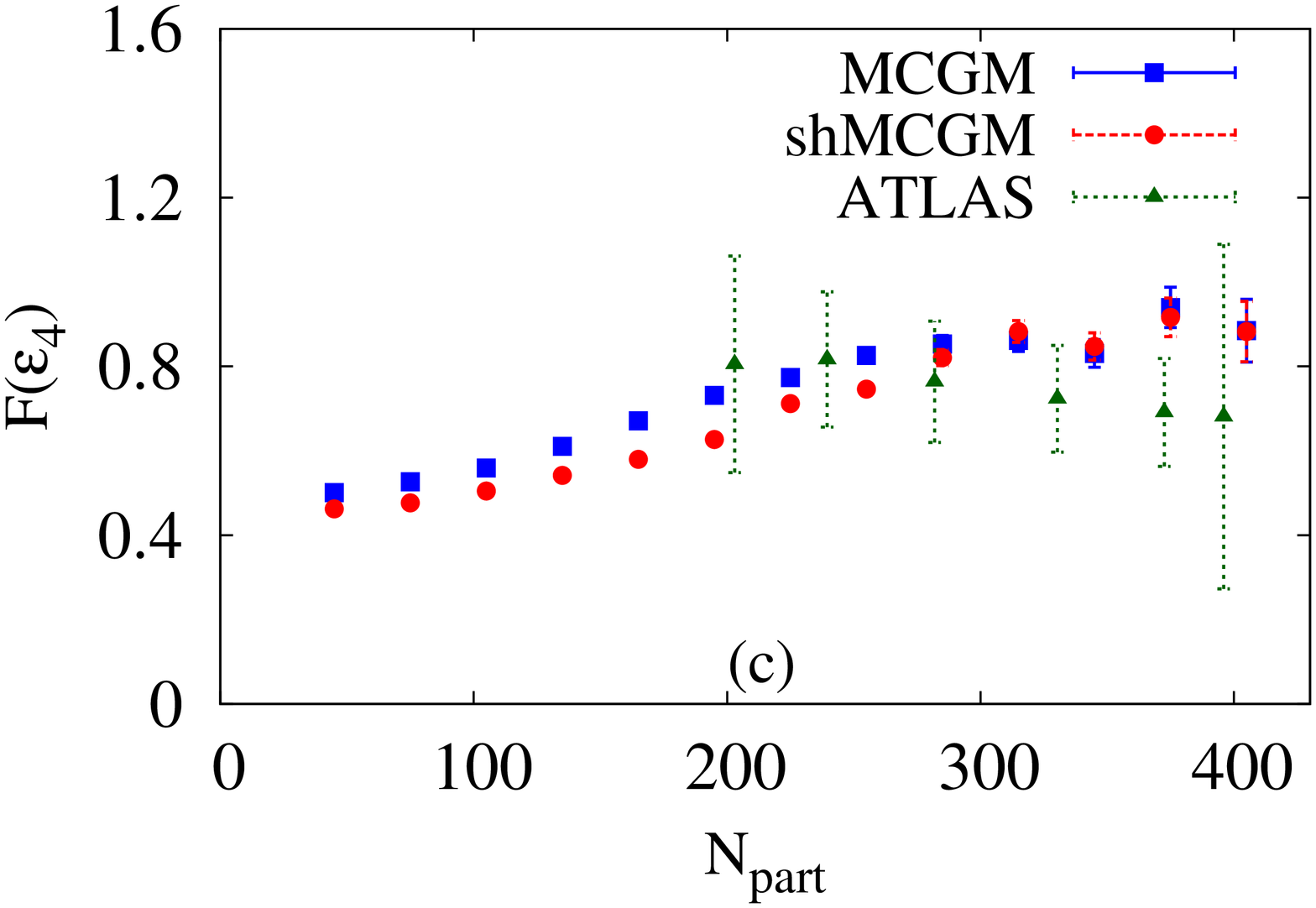}
\end{center}
\caption{(Color online) The centrality dependence of $F$ of the eccentricities compared 
between data~\cite{Aad:2014vba}, MCGM and shMCGM.}
\label{fig.Fen}\end{figure}

The standard deviation is a good measure of fluctuation about the mean value. ATLAS data is 
available on the scaled variance of the flow coefficients $v_n$ which can be compared to 
that of initial eccentricity assuming linear response~\cite{Aad:2013xma}. These quantities have been 
recently shown to agree well with MCGM computations~\cite{Rybczynski:2015wva}. In Fig.~\ref{fig.sigmaen}, we have 
plotted the centrality dependence of $\sigma\l\varepsilon_n\r/\la\varepsilon_n\ra$ in MCGM and shMCGM. We find 
almost equally good agreement between data and the different Glauber approaches. Two particle 
cumulants $\varepsilon_n\{2\}$ and four particle cumulants $\varepsilon_n\{4\}$ are other 
quantities of interest that are often measured in experiments and compared to theory,
\beqa
\varepsilon_n\{2\}&=&\la\varepsilon_n^2\ra^{1/2}\label{eq.en2}\\
\varepsilon_n\{4\}&=&\l2\la\varepsilon_n^2\ra^2-\la\varepsilon_n^4\ra\r^{1/4}\label{eq.en4}
\eeqa
From these cumulants it is possible to define the following scaled moments which 
within linear response will have one to one correspondence between the initial state(IS) and the 
final state (FS)
\beqa
F\l\varepsilon_n\r &=& \sqrt{\frac{\varepsilon_n\{2\}^2-\varepsilon_n\{4\}^2}
{\varepsilon_n\{2\}^2+\varepsilon_n\{4\}^2}}\label{eq.Fn}
\eeqa
We have plotted the centrality dependence of $F\l\varepsilon_n\r$ in Fig.~\ref{fig.Fen} and 
find good qualitative agreement between data~\cite{Aad:2014vba}, MCGM and shMCGM. Thus it is clear from 
Figs.~\ref{fig.e2e3},~\ref{fig.ebyee2},~\ref{fig.sigmaen} and \ref{fig.Fen} that the relative 
magnitudes of $\varepsilon_2-\varepsilon_3$ and E/E distribution plots of scaled $\varepsilon_2$ 
can discriminate clearly between MCGM and shMCGM. The rest of the observables have weaker 
discriminatory power.

\section{Summary and Discussions}\label{sec.summary}
Glauber models provide a simple and intuitive picture of the initial conditions for the system 
produced in heavy ion collisions. This model provides the centrality dependence of 
various ensemble average observables like charged particle multiplicity, anisotropies of the 
initial energy deposited etc. The MCGM generates event by event distributions of these 
quantities. The above observables including their 
event by event distribution can be measured in experiments and compared to such models of 
initial condition. Overall, Monte-Carlo Glauber models manage to provide a good 
qualitative description of the data. However, such geometrical models can not describe 
the recent data on $v_2-dN_{ch}/d\eta$ correlation from U+U at $\sNN=193$ GeV and event by event flow 
data for Pb+Pb at $\sNN=2.76$ TeV. This has called for a lot of ongoing effort to address 
such issues within the geometric approach of the Glauber model. We have earlier successfully 
addressed these issues at the top RHIC energy by introducing the effect of shadowing due to 
leading nucleons on the other participants located in the bulk~\cite{Chatterjee:2015aja}. Here 
we have extended our study to Pb+Pb collisions at $\sNN=2.76$ TeV. This idea of 
{\it eclipse} of nucleons in presence of other nucleons inside a nucleus has been proposed 
about sixty years back~\cite{Glauber:1955qq}. However, the significance of this effect in the 
phenomenology of heavy ion collisions has been hardly explored before. The current as 
well as our earlier work~\cite{Chatterjee:2015aja} suggest that the nucleon shadowing plays 
an important role in the ICs of HICs.

We find that for all centralities $\varepsilon_2$ is enhanced while $\varepsilon_3$ and $\varepsilon_4$ 
do not change much due to the inclusion of shadowing. This brings the predictions of the shadowed 
Glauber model for the mean values as well as event by event distributions of the eccentricities in 
agreement with IP-Glasma results. The relative magnitudes of 
$\varepsilon_2-\varepsilon_3$ and event by event distributions of $\varepsilon_2$ clearly demonstrate 
the superior performance of the shadowed Glauber model as compared to its conventional version, MCGM. The shadow 
parameter $\lambda$ drops by $\sim25\%$ as we go from top RHIC to LHC energies. This implies that at 
lower energies, the effect of shadowing in Glauber models will be even more important. Currently for 
low energies where the fireball is expected to carry a non-zero net baryon number, dynamical models 
are yet to be formulated and predictions made. On the other hand, within the ambit of the shadowed 
Monte-Carlo Glauber model, it is straightforward to make predictions at all energies as long as there 
is a good understanding of the $\sNN$ dependence 
of the model parameters.

{{\it Acknowledgement:}} We would like to thank Prithwish Tribedy for providing the IP-Glasma data. 
SC acknowledges him for many fruitful discussions on the initial condition and thanks ``Centre for 
Nuclear Theory" [PIC XII-R$\&$D-VEC-5.02.0500], Variable Energy Cyclotron Centre for support. SG 
acknowledges Department of Atomic Energy, Govt. of India for support. 
\bibliographystyle{apsrev4-1} 
\bibliography{LHCGlauberShadow}

\end{document}